\documentclass[reprint,aps,tightenlines,floatfix,showpacs]{article}
\usepackage{float}
\usepackage{graphicx}
\usepackage{bm}
\usepackage{amsmath}
\usepackage{amssymb}
\usepackage{mathrsfs}
\usepackage{paralist}
\usepackage{wrapfig}
\usepackage{placeins}
\usepackage{fancyhdr}
\newcommand{\cross}[2]{$ \{ #1 \} \, \otimes \; #2 $}
\newcommand{\scross}[2]{\begin{scriptstyle} \{ #1 \} \, \otimes \; #2 \end{scriptstyle}}
\newcommand{\sfrac}[2]{\begin{textstyle}\frac{#1}{#2}\end{textstyle}}

\newcommand{\ud}{\mathrm{d}}

\begin{document}
\bibliographystyle{unsrt}
\title{Fermion Confinement in Brane World Models with SO(10) Unification \\ \Huge{---} \\ \small{ \textit{The Rise and Fall of my Pet Universes}}\\ \Huge{---} }
\author{David Curtin \\ \\ \emph{School of Physics, University of Melbourne. Victoria 3010, Australia} \\ \\ Supervisor: Prof. Raymond Volkas}
\date{\today}

\begin{figure}
\begin{center}
\includegraphics{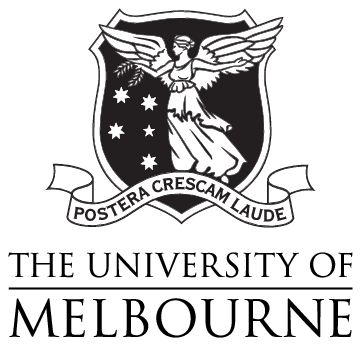}
\end{center}
\end{figure}

\maketitle
\thispagestyle{empty}

\begin{abstract}
In this Honours Research Project the aim is to build a toy model of
an SO(10) Grand Unified Theory with an extra spatial dimension.
Utilising the Clash of Symmetries mechanism proposed for brane-world
models by Davidson, Toner, Volkas and Wali and using the asymmetric
kink solution for the Higgs Field in the $45$ of SO(10) discovered
by Shin and Volkas, the symmetry on the brane will be broken to
$SU(4) \otimes SU(2) \otimes U(1)$, whereas the symmetry in the bulk
is $G_{SM} \otimes U(1)$. However, any fermions localised on the
brane would experience slight leakage into the bulk, and hence the
symmetry is broken down to $G_{SM} \otimes U(1)$. We then introduce
different SO(10) representations for fermions into the theory and
determine which components become localised as chiral zero-modes, in
the hope that the Standard Model particles are confined. Various
group theoretical techniques are introduced as we study the 120, 16,
126 and 144 representations, and detailed analysis of the 16's
confinement characteristics is performed.
\end{abstract}

\newpage
\thispagestyle{fancy}
\renewcommand\headrulewidth{0pt}
\lhead{\tiny{David Curtin}}
\rhead{\tiny{Honours Thesis, University of Melbourne, 2005}}
\cfoot{}
\subsection*{Statement of Contributions}
Sections \ref{sec:sm} and \ref{sec:background} as well as Appendices \ref{ap:gt} and \ref{ap:embeddings} are original reviews of the required background material. \\ \\
The group theoretical methods presented in Section \ref{sec:theory} and Appendix \ref{ap:spinordirectproduct} were developed by the author in the course of this research project, with the exception of Subsection \ref{subsec:howtobreakdownspinors}, which represents an original review of some of the basic theory available on Spinor Representations and includes much of the author's own derivations to ``fill in the blanks".\\ \\
The calculations and results presented in Section \ref{sec:fermions} and Appendices \ref{ap:calc120} and \ref{ap:16} are the author's own.

\subsection*{Acknowledgements}
I would like to thank my supervisor Ray, who always had time for me and was the best mentor imaginable, as well as my parents, whose continued love and support make my studies possible.

\newpage
\pagestyle{fancy}
\renewcommand\headrulewidth{0pt}
\lhead{\tiny{David Curtin}}
\rhead{\tiny{Honours Thesis, University of Melbourne, 2005}}
\cfoot{}
\tableofcontents
\newpage

\pagestyle{fancy}
\renewcommand\headrulewidth{0pt}
\lhead{\tiny{David Curtin}}
\rhead{\tiny{Honours Thesis, University of Melbourne, 2005}}
\cfoot{\thepage}
\setcounter{page}{1}

\section{Introduction}
\label{sec:intro}
The Standard Model (SM) of Particle Physics has very strong predictive powers and is an excellent model for most of physics as we know it. However, we know it cannot be the complete theory of how the universe works. There are several key problems:
\begin{compactitem}
\item[-] COBE and WMAP measurements of the cosmic microwave background indicate that the universe is about $\sfrac{3}{4}$ dark energy and $\sfrac{1}{5}$ dark matter, with the tiny rest being our usual SM baryonic matter. Dark matter is made up of particles which interact very weakly with our usual SM particles, including via gravity, and Dark Energy accelerates the expansion of the universe by counteracting gravity on large scales. The Standard Model supplies no suitable Dark Matter particles, and most quantum field theories predict a \emph{far} too large cosmological constant from the energy of the quantum vacuum.
\item[-] The SM does not include gravity. Gravity is described by General Relativity which breaks down for extremely small distance scales and high mass densities. Clearly a theory of quantum gravity is needed if we are to attain a complete picture of the universe's particles and interactions.
\item[-] The SM contains unpleasantly many free parameters -- about 20, more with right-handed neutrinos. Things like gauge coupling constants and fermion masses have to be put in by hand, i.e. determined from experiments. It would be more satisfying to derive the values of these parameters from higher principles.
\item[-] In order to properly adjust the Higgs mass (and ultimately all other particle masses), we have to adjust the difference between the squares of the self- and fermion-coupling constants of the Higgs field to within one part in $\sim 10^{15}$. If we assigned no physical significance to the SM cut-off parameter $\Lambda$, this fine-tuning would not be a problem. However, $\Lambda$ is the energy scale of quantum gravity at which we expect the three gauge forces to unite and a new, non-SM physics to emerge. This fine-tuning is known as the ``hierarchy problem" \cite{abdel}. 
\item[-] Our universe today is composed of ordinary matter with almost no antimatter. There was a small matter-excess early in the universe which leads to our existence, and we have no explanation for it. 
\item[-] The SM in no way explains \emph{why} there are three separate gauge forces and three copies of all the fermions (three generations)
\end{compactitem}
There are several possible angles of attack for extending the standard model. Often, to try out new possibilities, it is useful to deal with toy models which only address a few components of a full theory. If the results are promising, further work to make the model more realistic is justified.

In this Honours Research Project, the aim is to build a toy model of an SO(10) Grand Unified Theory with an extra spatial dimension. Building on previous results in the field, we will try out different SO(10) representations for the fermions and attempt to localise the SM particles on branes in order to reproduce our 4-dimensional world. In doing so, many important group theoretical techniques are developed. The favourite candidate is the 16 since it perfectly contains just the SM particles. As expected its preliminary results are the most promising, and we perform a detailed analysis of its confinement characteristics. While the final results are not as immediately appealing as we wished, we remain hopeful that it may be possible to develop a suitable model using a more general solution of the Higgs Potential. At any rate, the group theoretical machinery developed during this project will be useful to anyone doing model building of this type in the future.

The outline of this thesis is as follows. In Section \ref{sec:sm} we will provide a quick review of the main concepts of the Standard Model. In Section \ref{sec:background} the fairly extensive amount of background theory for this project is introduced. Section \ref{sec:theory} develops the theoretical techniques required to perform the necessary calculations, which are then presented in Section \ref{sec:fermions}.

\section{Quick Review of the Standard Model}
\label{sec:sm}
It is impossible to give a comprehensive introduction to the SM for the completely uninitiated in the limited space available here. Rather, this Section will serve as a reminder of the main concepts important in this research project. Much of the material here follows \cite{cheng}. An introduction to the necessary group theory is given in Appendix \ref{ap:gt}\footnote{Note that this Section stands separately and does not necessarily employ exactly the same naming or sign conventions. Any differences however are completely insubstantial.}

\subsection{Gauge Symmetries}
Consider some fermions $\Psi_{1,\alpha}, \Psi_{2,\alpha},\ldots ,\Psi_{N,\alpha}$, each of which are independent 4-D Lorentz Spinors\footnote{Roman letters denote SU(N) indices, Greek letters denote Lorentz/Spinor indices.}. Let them form an $N$ representation of SU(N).

\begin{equation*}
\Psi = \left( \begin{array}{c} \Psi_1\\ \vdots \\ \Psi_N  \end{array} \right).
\end{equation*}
To simplify things we will often use matrix/vector notation to sum over SU(N) indices, i.e. write $U \Psi$ for $U^j_a \Psi^a $, $\boldsymbol{\tau \cdot \theta}$ for $\tau^a \theta^a$ etc. Often we will also suppress the Spinor indices of the $\Psi$'s when they are of no immediate interest.

The free Lagrangian Density of these fermions is
\begin{equation*}
\mathscr{L} = \overline{\Psi}(x)(i \gamma^\nu \partial_\nu - m)\Psi(x).
\end{equation*}
This Lagrangian has a \emph{global} SU(N) symmetry manifested by invariance under the transformation $U(\boldsymbol \theta) = e^{-i \boldsymbol{\theta \cdot \tau}}$, where $\theta^a$'s are the group parameters and the $\tau^a$'s the generators of SU(N) satisfying $[\tau^i,\tau^j] = i f^{ijk} \tau^k$. In this case global means that the parameters of the transformation are independent of $x^\mu$. Two things to note: 
\begin{compactenum}
\item If N = 1, the symmetry is U(1) and there is only one generator, obviously implying $f^{ijk} = 0$. In that case the analysis following simplifies somewhat.
\item The mass term is only invariant if the right- and left-handed components of each fermion belong to the same representation (more on that later).
\end{compactenum}
Now generalise to a \emph{local} or \emph{gauge} symmetry, i.e. allow the group parameters to vary with position in space-time: 
\begin{equation*}
U(\boldsymbol{\theta}(x^\mu)) = e^{-i \boldsymbol{\theta}(x^\mu)\cdot \boldsymbol{\tau}}.
\end{equation*}
Due to the space-time derivative, $\mathscr{L}$ is no longer invariant (keep in mind that the SU(N) transformation and the $\gamma$-matrices commute since they act on different spaces):
\begin{align}
\overline{\Psi}(x^\mu)\partial_\nu \Psi(x^\mu) \longrightarrow &[\overline{\Psi}(x^\mu) {U^\dagger}(\theta)] \partial_\nu [U(\theta)\Psi(x^\mu )] \nonumber \\
= &  \overline{\Psi}(x^\mu)\partial_\nu \Psi(x^\mu) + \overline{\Psi}(x^\mu){U^\dagger}(\theta)[\partial_\nu U(\theta)] \Psi(x^\mu). \nonumber
\end{align}
Clearly the last term is not zero for a local transformation. To make the Lagrangian invariant again, we replace the regular derivative by the \emph{gauge-covariant derivative}:
\begin{equation*}
D_\mu = \partial_\mu - i g \boldsymbol\tau \cdot \boldsymbol{A}_\mu.
\end{equation*}
Here we have introduced the vector gauge fields $A_\mu^j$, one for each generator. They act as force carriers, coupling to the fermions with strength g via the second term. The photon, for example, is the $U(1)_{EM}$ gauge boson, and two charged particles interact with each other by exchanging virtual photons, where the coupling strength of the photon/fermion interaction is the fermion's electric charge.

The condition that $\mathscr{L}$ be invariant under a local gauge transformation imposes certain transformation properties onto the $A_\mu^j$'s. We require that the covariant derivative of $\Psi$ transforms like $\Psi$ itself, i.e.
\begin{equation*}
D_\mu \Psi \rightarrow U(\boldsymbol \theta) D_\mu \Psi
\end{equation*}
which requires the gauge fields to transform as
\begin{equation} \label{eq:gaugefieldtransform}
\boldsymbol \tau \cdot \boldsymbol{A}_\mu \longrightarrow U(\boldsymbol \theta) \boldsymbol \tau \cdot \boldsymbol{A}_\mu {U^\dagger}(\boldsymbol \theta) - \sfrac{i}{g}[\partial_\mu U(\boldsymbol \theta)] {U^\dagger}(\boldsymbol \theta)
\end{equation}
The representation of the gauge fields becomes apparent when looking at the infinitesimal transformation $U = 1 - i \boldsymbol{\tau \cdot \theta}$. In this case, Equation \ref{eq:gaugefieldtransform} reduces to
\begin{equation*}
A^i_\mu \longrightarrow A^i_\mu + f^{ijk}\theta^j A^k - \sfrac{1}{g} \partial_\mu A^i
\end{equation*}
where we recognise that the second term is the transformation for the adjoint representation of SU(N). Hence, for $N > 1$, the $A^i_\mu$'s form an adjoint representation and carry gauge charges.

In order to make the gauge fields dynamical variables we have to introduce their kinetic term into the Lagrangian. The simplest gauge-invariant term of this form (conventionally normalised) is
\begin{equation*}
-\sfrac{1}{4} F^a_{\mu \nu} F^{\mu \nu}_a
\end{equation*}
where
\begin{equation*}
F^a_{\mu \nu} = \partial_\mu A^a_\nu - \partial_\nu A^a_\mu + g f^{abc} A^b_\mu A^c_\nu
\end{equation*}
Hence the final gauge-invariant Lagrangian is
\begin{equation*}
\mathscr{L} = -\sfrac{1}{4} F^a_{\mu \nu} F^{\mu \nu}_a + \overline{\Psi}(x)(i \gamma^\nu D_\nu - m)\Psi(x).
\end{equation*}

\subsection{The Higgs Mechanism}
The Gauge Group of the SM is $SU(3)_c \otimes SU(2)_W \otimes U(1)_Y$, where c, W, and Y stand for colour (strong force), weak and hypercharge respectively. Hence in building up the SM Lagrangian we include in our covariant derivative gauge fields and couplings for each of the three simple groups, as well as three gauge field kinetic terms. Each of the SM particles is part of a representation of that gauge group:
\begin{equation}
\label{eq:smreps}
\begin{array}{ll}
f_{iL} = {\left(\begin{array}{c} \nu_i \\ e_i \end{array} \right)}_L \thicksim (1,2)(-1) \qquad \qquad& 
Q_{iL} = {\left( \begin{array}{c} u_i \\ d_i \end{array} \right)}_L \thicksim  (3,2)(\sfrac{1}{3}) \\
e_{iR} \phantom{= {\left( \begin{array}{c} \nu_i \\ e_i \end{array} \right)}_L} \, \, \thicksim (1,1)(-2) \qquad \qquad& 
u_{iR} \phantom{= {\left( \begin{array}{c} \nu_i \\ e_i \end{array} \right)}_L}\; \;  \thicksim (3,1)(\sfrac{4}{3}) \\
\nu_{iR} \phantom{= {\left( \begin{array}{c} \nu_i \\ e_i \end{array} \right)}_L} \, \, \thicksim (1,1)(0) \qquad \qquad& 
d_{iR} \phantom{= {\left( \begin{array}{c} \nu_i \\ e_i \end{array} \right)}_L}\; \;  \thicksim (3,1)(-\sfrac{2}{3})
\end{array}
\end{equation}
where, for example, $(1,2)(-1)$ means that the left-handed electron and neutrino together transform as a colour (SU(3)) singlet, form a weak (SU(2)) doublet and have (U(1)) hypercharge $-1$. The index $i$ is not a group theory index here but runs from 1 to 3 to label the three generations of fermions, and the left/right-handed components of each Lorentz fermion are projected out using $\sfrac{1}{2}(1 \mp \gamma^5)$, where $\gamma^5 = i\gamma^0 \gamma^1 \gamma^2 \gamma^3$ is the standard 4-D chirality operator.

It is clear that fermion mass terms are as such not gauge invariant:
\begin{equation*}
m_e \overline{e}_L e_R \thicksim (1,2)(-1)
\end{equation*}
(bear in mind that conjugating the fermion conjugates its representation). We also have to explain the massive bosons mediating weak interaction, which is not accounted for using only gauge symmetries. We will therefore introduce a Higgs boson in an appropriate potential and use \emph{spontaneous symmetry breaking}, which will break the gauge symmetry from $SU(3)_c \otimes SU(2)_W \otimes U(1)_Y \longrightarrow SU(3)_c \otimes U(1)_Q$ where $U(1)_Q$ is the normal electromagnetic gauge group \footnote{EM charge is given as $Q = (I_{3L} + \sfrac{Y}{2})$, where Y is the hypercharge and $I_{3L}$ is the appropriate entry of the diagonal generator of the SU(2) rep the particle belongs to.}.

In order for the fermion mass terms to be singlets under the gauge group, we can give them masses via a Yukawa interaction with the Higgs field $\phi$
\begin{equation*}
\lambda \overline{e}_L e_R \phi + \textrm{ hermitian conjugate}
\end{equation*}
which requires that $\phi \thicksim (1,2)(1)$, i.e. a weak complex doublet scalar field\footnote{since for SU(2), $2 \otimes 2 = 1 \oplus 3$ and the singlet gives us our mass term}. We write
\begin{equation*}
\phi = \left( \begin{array}{c} \phi^+ \\ \phi^\circ \end{array} \right).
\end{equation*}
Note that $\phi$ has 4 real components. The electron mass term  becomes
\begin{equation*}
\mathscr{L}_{\textrm{mass}} = \lambda (\overline{\nu}, \overline{e})_L \left( \begin{array}{c} \phi^+ \\ \phi^\circ \end{array} \right) e_R + \textrm{ h.c.}
\end{equation*}
which is gauge invariant (similarly for quarks). Obviously this only works if $\phi$ takes on an appropriate vacuum expectation value (vev). Insert the potential $V(\phi) = \lambda (\phi^\dagger \phi - u^2)^2$ into the Lagrangian, where $\lambda > 0$ so the potential is bounded from below. This potential is SU(2) invariant and the vacuum (i.e. the lowest-energy state) only requires that $<\phi^\dagger \phi>_\circ = u^2$. 

We now \emph{break the symmetry} by fixing the vacuum to be some distinct value (a more thorough review of spontaneous symmetry breaking is given in \ref{subsec:ssb}). This is really just what nature does -- the Lagrangian might be invariant under some symmetry but the Higgs vacuum invariably chooses one of the possible vacua and stays there, breaking that symmetry. In order to get the correct form of the fermion mass term, we can always use an SU(2) rotation to change our basis such that the vacuum state of the Higgs is
\begin{equation*}
<\phi> = \left( \begin{array}{c} 0 \\ u \end{array} \right),
\end{equation*}
i.e. $<\textrm{Re}(\phi^\circ)>_\circ = u$ whereas the three other components -- called the Goldstone Bosons -- have zero vev's. The weak boson mass terms are then produced by the gauge covariant derivative (the Higgs kinetic term) of the Higgs. Expanding
\begin{equation*}
\mathscr{L}_{\textrm{Higgs}} = (D_\mu \phi)^\dagger (D^\mu \phi) + \ldots
\end{equation*}
around the vacuum and redefining our gauge bosons, we get mass terms for the W and Z bosons, whereas the photon stays massless\footnote{The photon is represented by a linear combination of the $U(1)_Y$ gauge boson and an SU(2) gauge boson and stays massless. The orthogonal linear combination (Z) together with the two remaining SU(2) bosons (W's) gain mass and mediate the weak force. The gluon stays massless since the $\phi$ is a colour singlet.}. The three goldstone bosons have actually been ``eaten up", i.e. incorporated into the three massive gauge bosons, giving them the extra degree of freedom (longitudinal polarisation) required to be massive. Since the mass terms are proportional to $u$, the Higgs vev sets the scale for the weak boson masses. Similarly, by substituting the Higgs vacuum value into the fermion Yukawa coupling term we get our fermion masses. This is called the \emph{Higgs Mechanism}.

We now have the complete Standard Model. The SM Lagrangian is made up of the Higgs terms to supply mass to appropriate gauge bosons and fermions (covariant derivative kinetic term, potential $V$ and Yukawa coupling to fermions), the three kinetic terms for the gauge fields, as well as the fermion kinetic term, which in its covariant derivative includes interaction terms with the gauge bosons. Note that the Standard Model is \emph{chiral}, i.e. states of different chirality come in different representations and hence interact differently with respect to the weak force.

\section{Extending the Standard Model}
\label{sec:background}
There are many ways to try and extend the Standard Model of particle physics. The following Sections will give a brief overview of those utilised by my research project.

\subsection{Grand Unified Theories}
\label{subsec:gut}
The gauge group of the standard model is $G_{SM} = SU(3)_c \otimes SU(2)_W \otimes U(1)_Y$. Quantum chromodynamics with $SU(3)$ is the theory of the strong force, and the Glashow-Weinberg-Salam model with $SU(2) \otimes U(1)$ provides the theory of weak and electromagnetic interactions. 

Group theoretically, it would clearly be desirable to unify all three interactions into one simple gauge group, which at some high scale $M_X$ is broken down to the standard model. Possible $G_{GUT}$ candidates obviously have to satisfy $G_{GUT} \supset G_{SM}$. We also have to work with representations of $G_{GUT}$ which, upon restriction to $G_{SM}$, break down to yield the $SU(3) \otimes SU(2) \otimes U(1)$ representations of the standard model fermions (see Equation \ref{eq:smreps}). Of course other particles might also be produced, and that may even be desirable if one wanted to explain for example Dark Matter. 

Often the GUT includes a Higgs field in the adjoint representation, which allows us to break a group down to maximal subgroups. This brings us to one of the key feature of these models, and also their greatest weakness. They require intermediate stages of symmetry breaking in order to make them resemble reality, so Model builders, having just come up with these beautiful gauge symmetries, now have to introduce additional Higgs fields into the theory in order to perform the breakdown. This in turn requires extra parameters for couplings, masses etc, so instead of reducing the number of  degrees of freedom we find ourselves with more than when we started. However, there are ways of addressing this issue, one of which is the Clash of Symmetries Mechanism, which will be discussed in \ref{subsec:ssb}.

A good first example of a Grand Unification Group is  $SU(5)$, proposed by Georgi and Glashow in 1973 \cite{gut}. $SU(5)$ has $G_{SM}$ as a maximal subgroup, and under the restriction $SU(5) \supset SU(3) \otimes SU(2) \otimes U(1)$ its $5$ and $\overline{10}$ break down quite nicely to accommodate the SM particles.
\begin{equation*}
5 \longrightarrow \underbrace{(3,1)(-\sfrac{2}{3})}_{d_R} \oplus \underbrace{(1,2)(1)}_{(f_L)^c} \qquad \qquad 10 \longrightarrow \underbrace{(\bar{3},1)(-\sfrac{4}{3})}_{(u_R)^c} \oplus \underbrace{(1,1)(2)}_{(e_R)^c} \oplus \underbrace{(3,2)(\sfrac{1}{3})}_{Q_L}
\end{equation*}
where the superscript $c$ stands for charge conjugation, which is nothing but a basis change to antiparticle basis that has the effect of conjugating the corresponding representation.
The $SU(5)$ gauge bosons are in the $24$, the adjoint representation, which breaks down as follows:
\begin{equation*}
24 \longrightarrow \underbrace{(8,1)(0)}_{\textrm{gluons}} \oplus \underbrace{(1,3)(0)}_{W^\mu_i} \oplus \underbrace{(1,1)(0)}_{B^\mu} \oplus \underbrace{(3,2)(-\sfrac{5}{3})}_{X,Y} \oplus \underbrace{(\bar{3},2)(\sfrac{5}{3})}_{\overline{X}, \overline{Y}}
\end{equation*}
Apart from the SM force carriers, there are some new gauge bosons
\begin{equation*}
\left( \begin{array}{c} X^{-\frac{1}{3}} \\ Y^{-\frac{4}{3}} \end{array} \right) \thicksim (3,2)(-\sfrac{5}{3})
\end{equation*}
(superscript denotes electric charge, colour freedom has been suppressed). 

There are two Higgs fields in the traditional formulation of this theory: an SU(5) $5$ which contains a weak doublet to break the electroweak symmetry, and an adjoint $24$, which breaks $SU(5)$ to $G_{SM}$ at a scale $M_X$, giving the X and Y bosons a mass at the $M_X$ scale. Since we don't see these exotic gauge bosons in everyday physics, $M_X$ has to be very large. These bosons are interesting because they can mediate processes which violate baryon/lepton number, maybe pointing to a solution to the matter-excess problem. However that also makes them troublesome since they can mediate proton decay. The measured lower bound for proton lifetime of $\gtrsim 10^{30}$ years requires that $M_X\gtrsim 10^{15}$ GeV.

Other potential GUT candidates include SO(10) \cite{gut} and $E_6$. SO(10) Grand Unification, utilised in this project, is of great interest since the $16$ of SO(10), under restriction to SU(5), breaks down to $10 \oplus \bar{5} \oplus 1$, where the singlet can represent the right-handed neutrino.

\subsection{Extra Dimensions}
\label{subsec:extradim}

\subsubsection{Overview}

Introducing extra dimensions was first attempted in the 1920's by Kaluza and Klein. Trying to unify electromagnetism with gravity, they introduced a compact $5^{\textrm{th}}$ dimension and assumed that the photon originated from the new components of the now 5-dimensional metric tensor. Today we can try to use extra dimensions in our theories to address several of the Standard Model's shortcomings. Sometimes it is even required -- String Theory, a candidate for a theory of quantum gravity, is only consistent in 10 or 11 dimensions, since otherwise malignant ghost quantum states with unphysical negative probabilities spoil Poincar\'e (i.e. translational) invariance.

Let us first define two key concepts important to our discussion of extra dimensions: \emph{compactification} and \emph{branes}. 
Compactification simply means that the extra dimension(s) are rolled up such that they form a compact space, i.e. not infinite (unlike, it would appear, our standard dimensions). 
The word \emph{brane} is a short form of membrane, and brane-world models postulate that our universe is confined to a hypersurface embedded in a higher-dimensional space, called the \emph{bulk}. Addressing the issue of confinement is done in one of two ways. One can simply assume this brane, with the SM fields living on it, to exist \emph{a priori}, or one can generate it dynamically by finding some confinement mechanism which stops the SM fields from propagating through the bulk in the low-energy regime. A dynamic confinement mechanism important to this project is discussed later in this Section.

Broadly speaking, we can classify models with extra dimensions into three classes: \\ \\
\textbf{Small compactified extra dimensions}\\
Working only with compactified extra dimensions was the standard approach for some time. In such a model the world appears 3+1 dimensional as long as we only look at distance scales larger than the compactification radius (to a tightrope walker the rope is an essentially one-dimensional object, but an ant crawling on it sees all 3 dimensions). In order to avoid contradiction with experiment the confinement radius has to be extremely small, $\lesssim 10^{-18}$m. Initially this is how String Theorists hid their additional dimensions, but in recent times there has been a shift to consider brane-world models also, which can be dynamically generated in String Theory. \\ \\
\textbf{Compactified extra dimensions with branes}\\
A motivation for relatively large extra compactified dimensions is a possible solution to the hierarchy problem: the fundamental scale would not be the Planck Scale $M_P \sim 10^{18} GeV$, but rather its 5-dimensional equivalent $M_*$. Working in natural units, they are related via $M_P^2 = M_*^{n+2} V_n$, where $V_n$ is the (large) volume of the extra space. The hierarchy problem ceases to be one, since the smaller fundamental scale requires no fine tuning to cancel out divergences. The ADD model \cite{add} falls in this class. Arkani-Hamed, Dimopoulos and Dvali proposed in 1998 that we can have compactified extra dimensions far larger than the Planck scale if we confine the SM fields to a brane-like object. Only gravity, being an integral part of space-time, is allowed to propagate in the bulk. The SM fields are confined to the brane by trapping fermionic zero modes at topological defects (similar to the mechanism in \ref{subsubsec:dynamicconfinementmechanisms}), and the graviton's propagation in the bulk has measurable consequences, since it changes the effective 4-dimensional gravitational potential below a certain size scale, which for two additional dimensions is in the sub-mm range. \\ \\
\textbf{Infinite extra dimensions} \\
We can have infinite extra dimensions if we confine our universe to a brane. While this gives us useful model building freedom, the confinement of gravity presented a problem, since an infinite extra dimension gives rise to a continuous Kaluza Klein mass-spectrum for the graviton\footnote{The spectrum of higher-mass modes of a particle allowed to propagate in the bulk, unlike the confined massless zero-mode}, meaning that low-energy modes are accessible and change the behaviour of gravity in 4 dimensions.  This problem is addressed by the RS2 model, developed by Randall and Sundrum in 1999 \cite{rs2}, where the additional dimension is curved. Through use of a certain non-factorisable background-metric, the massless graviton is confined to a single 3-brane\footnote{i.e. 3 spatial dimensions} with positive tension, and reproduces 4-dimensional gravity. The massive KK modes only contribute an exponentially small correction. This method of localising gravity is extremely useful since it can be used in many kinds of brane-world models with infinite extra dimensions, opening up the possibility of using them to extend the Standard Model.

\subsubsection{A Dynamic Fermion Confinement Mechanism}
\label{subsubsec:dynamicconfinementmechanisms}

We will now outline a fermion confinement mechanism which was discovered by Jackiw and Rebbi in 1976 \cite{rebbi} and was first applied to dynamically create branes by Rubakov and Shaposhnikov in 1983 \cite{originalbranes}. Consider the free five dimensional Dirac Equation for a fermion $\Psi$. Introduce a Yukawa coupling to a scalar $\phi$ which changes only in the 5th dimension
\begin{equation} \label{eq:simpledirac}
0 = i \Gamma^K\partial_K\Psi + g \phi(z) \Psi.
\end{equation}
where $K$ is a five-dimensional Lorentz index and $\{\Gamma^\mu = \gamma^\mu, \Gamma^5 = -i\gamma^5\}$ are the gamma matrices of the 5-D Clifford Algebra.

Let $\phi$ be an odd function\footnote{can be used to represent a topological defect} which passes through zero at $z_0$. One example would be $\phi = \tanh{\mu z}$ with $z_0 = 0$. Then there exists a fermionic zero-mode confined at $z = z_0$. To see this, let $\Psi$ be 4-dimensional zero-mode and write it in terms of left and right chiral components: 
\begin{equation}\label{eq:psiLR}
\Psi(x^\mu,z) = {f_L}(z) \Psi_{L0}(x^\mu) + {f_R}(z) \Psi_{R0}(x^\mu) \qquad \textrm{where} \qquad \gamma^\mu \partial_{\mu} \Psi_{L,R0} = 0
\end{equation}
With this separation of variables we can now solve Equation \ref{eq:simpledirac} for left- and right-handed components separately. Deal with the left-handed component first:
\begin{equation*}
0 = i \Gamma^K \partial_K \left[ f_L(z) \Psi_{L0}(x^\mu) \right] + g \phi(z) \left[ f_L(z) \Psi_{L0}(x^\mu) \right]
\end{equation*}
Using equation \ref{eq:psiLR} this reduces to 
\begin{equation*}
0 = i(\Gamma^5 \Psi_{L0})(\partial_5 f_L) + g \phi f_L \Psi_{L0}
\end{equation*}
$\Psi_{L0}$ is a chiral eigenstate: $i \Gamma^5 \Psi_{L0} = \gamma^5 \Psi_{L0} = -\Psi_{L0}$, so we can divide out $\Psi_{L0}$ and obtain a simple differential Equation for $f$:
\begin{equation*}
\partial_z {f_L}(z) = g{\phi}(z) {f_L}(z)
\end{equation*}
which has the solution
\begin{equation} \label{eq:fLsolution}
f_L(z) \propto e^{ \int_{z_0}^z g \phi (z') \ud {z'}}.
\end{equation}
Similarly, the right-handed Equation yields
\begin{equation} \label{eq:fRsolution}
f_R(z) \propto e^{- \int_{z_0}^z g \phi (z') \ud {z'}}.
\end{equation}
So if $g \phi(z) <$ or $> 0$ for $z > z_0$, the confined component of the fermionic zero mode is left- or right-handed respectively. Note that these confined fermions are massless 4-dimensional Weyl Spinors, and in realistic models would have to be given a small mass by some other mechanism.

This fermion confinement mechanism will be at the centre of this research project. The general issue of confinement is not completely resolved -- while the RS2 model offers a nice way of confining gravity, the so far proposed mechanisms for confining gauge fields are less compelling, and work in that area is still in progress.

\subsection{Kinks and Spontaneous Symmetry Breaking Mechanisms}
\label{subsec:ssb}
Just because the Lagrangian of a system displays a certain symmetry does not automatically imply that this symmetry is also present in the ground state. For example, the Lagrangian of a ferromagnet is rotationally invariant, whereas any actual ferromagnet below critical temperature is not. We can conjecture that the laws of nature display symmetries which may not be manifest at or close to the ground state. 

In this Section we will outline the standard \emph{Spontaneous Symmetry Breaking} (SSB) mechanism, already utilised in Section \ref{sec:sm} for the Higgs Mechanism, as well as an enhanced version called \emph{Clash of Symmetries} (COS), in particular how it relates to this project.

\subsubsection{Spontaneous Symmetry Breaking}
The basic idea behind Spontaneous Symmetry Breaking (SSB) is as follows. Consider a Lagrangian Density which is invariant under transformations belonging to some symmetry $G$. We can introduce a Higgs Field $\phi$, which has certain transformation properties under $G$, by inserting its kinetic term as well as a potential V:
\begin{equation*}
\mathscr{L} = (D_\mu \phi)^\dagger (D^\mu \phi) - V(\phi) + \textrm{ \{possibly other terms, e.g. gauge field kinetic terms\},}
\end{equation*}
where both the (possibly covariant) derivative term and the potential are invariant under $G$. The \emph{vacuum manifold} is the configuration space for the Higgs representing \emph{minima} of the potential. Looking at any one point on the vacuum manifold, any symmetry transformation belonging to $G$ can be classified in one of two ways: either it leaves the point invariant, or it moves it to a different location on the vacuum manifold. The latter symmetry transformations \emph{generate} the vacuum manifold (i.e. using them we can connect any two points on the vacuum manifold). 

In the physical system described by this Lagrangian, invariably the Higgs Field (which generally is a function of space-time coordinates) assumes some definite configuration. At the boundary points (usually at spatial infinities) the Higgs assumes definite values on the vacuum manifold. In doing so, it has \emph{spontaneously broken} the symmetry which generates the manifold, and hence the original symmetry $G$ has been reduced to $H$, the \emph{unbroken} symmetry which leaves the vacuum invariant. 

This mechanism is often used to break gauge symmetries down to some suitable subgroup. If we examine the SSB mechanism in the context of Lie groups, we see that those Lie Group generators annihilating the Higgs vacuum generate transformations which leave it invariant. Let there be $N$ total generators, and $M$ which annihilate the vacuum. This means that there are $N-M$ group generators generating the transformations belonging to the coset space $G/H$, which define the vacuum manifold. These generators are associated with $N-M$ massless \emph{Goldstone Bosons}, which are, upon SSB, ``eaten up" by gauge fields to make the gauge bosons massive. As was explained in Section \ref{sec:sm}, we can employ a $G_{SM}$ $(1,2)(1)$ Higgs multiplet to break the symmetry down to $SU(3)_c \otimes U(1)_Q$ and make the weak bosons massive (as well as the fermions, if we introduce the required Yukawa coupling).

\subsubsection{The Clash of Symmetries Mechanism}

As was pointed out in \ref{subsec:gut}, one problem with Grand Unified Theories is that the beautifully enhanced symmetries have to be broken down in various stages by different Higgs Fields, each of which requires a large number of free parameters.  This is because, when working with Higgs fields in 3+1 dimensions, we are restricted to using Higgs vacuum solutions that are constant in space as well as time, since within our Hubble Volume there is no evidence for topological defects characteristic of soliton solutions (i.e. solutions which are stable but non-uniform).  If we could use soliton solutions, the ``symmetry breaking power" of a Higgs could be greatly enhanced. This is the motivation of the \emph{Clash of Symmetries} (COS) mechanism, developed by Davidson, Toner, Volkas and Wali \cite{volkascos}. \\
\begin{wrapfigure}{l}{4cm}
\includegraphics[keepaspectratio, width = 4cm]{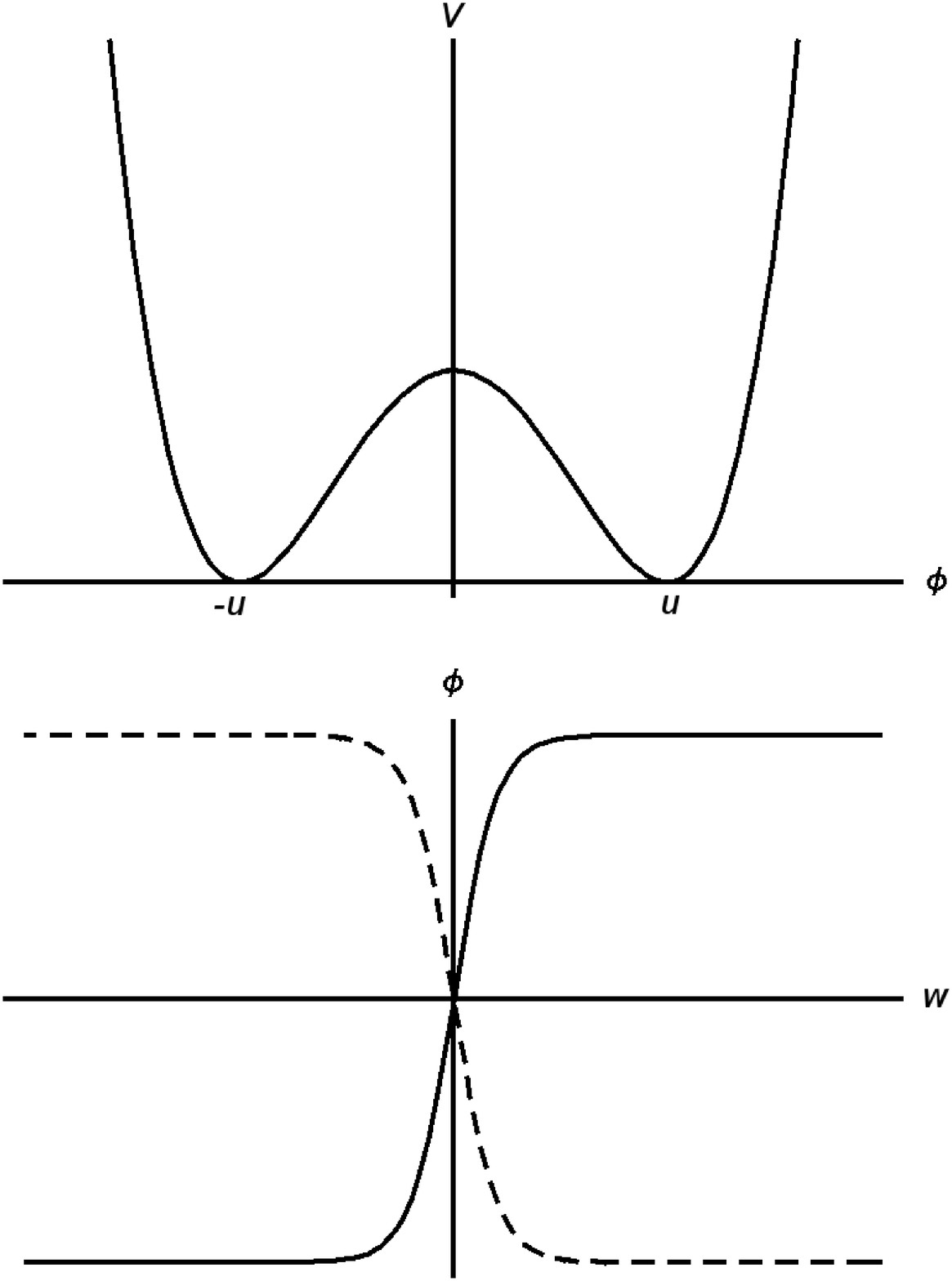}
\caption{The $\mathbb{Z}_2$ symmetric potential and kink solutions}
\label{fig:z2kinkstuff}
\end{wrapfigure}
\textbf{The $\mathbb{Z}_2$ Kink}\\
For a start, consider the simple $\mathbb{Z}_2$ kink. The potential for the Higgs is $V = \lambda(\phi^2 - u^2)^2$ which has the discrete $\mathbb{Z}_2$ symmetry $\phi \rightarrow -\phi$. The vacuum manifold is made up of just two points. The degenerate global minima are $\phi = \pm u$, which are clearly by themselves solutions to the Euler-Lagrange Equations for the Higgs. \emph{Now allow $\phi$ to vary along one dimension $w$}. The $\mathbb{Z}_2$ kink/antikink is a static solution to the Euler Lagrange Equations which interpolates between the global minima of $V$ and has the form $\phi(w) = \pm u \tanh{(\sqrt{2 \lambda} u w)}$. While these kinks have higher energy density than the constant vacuum solution $\phi = \pm u$, they are stable since the global minima defining the boundary conditions are disconnected from each other. The domain wall at $w = 0$ is brane-like. 

\begin{wrapfigure}{r}{5.5cm}
\includegraphics[keepaspectratio, width = 5.5cm]{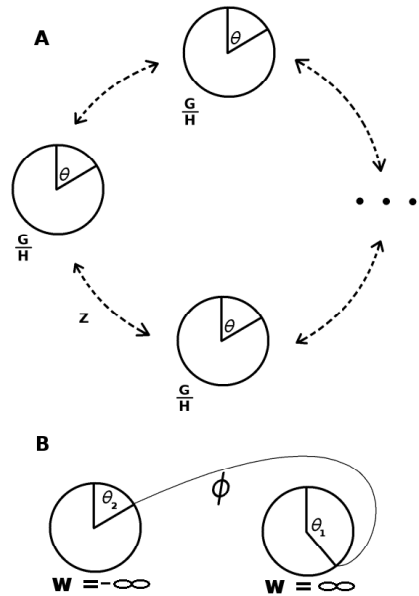}
\caption{\textbf{A}: The disconnected vacuum manifold. \textbf{B}: A possible soliton Higgs solution.}
\label{fig:cos}
\end{wrapfigure}

The lesson to draw from this simple kink example is: if the vacuum manifold is made up of disconnected points related to each other via some discrete symmetry $Z$, we can break the $Z$ symmetry and get soliton solutions interpolating between different points on the vacuum manifold. \\ \\
\textbf{Clash of Symmetries}\\
Now let us introduce an additional continuous symmetry $G$ into the Lagrangian, which is broken down to $H$. Then \emph{each} of the points on the vacuum manifold, related to each other via $Z$, ``expands" into a sub-manifold generated by a copy of the coset space $G/H$. In Figure \ref{fig:cos}A, the circles represent the sub-manifolds. Crucial to the Clash of Symmetries Mechanism is the fact that there are several possible ways to embed $H$ in $G$, schematically represented by the angle $\theta$.

We can now obtain some highly interesting soliton solutions. Again, for simplicity, say that the Higgs $\phi$ only varies in one dimension $w$. To get a soliton, let the Higgs interpolate between the vacua on two different sub-manifolds (anologous to interpolating between the two different points of the vacuum manifold for the $\mathbb{Z}_2$ kink). 
We can set the boundary conditions such that the $H$ symmetry of the first sub-manifold is embedded differently to the $H$ symmetry on the second sub-manifold. In Figure \ref{fig:cos}B, this is represented by different choices for the two $\theta$'s. Now keep $\theta_2$ constant while changing $\theta_1$ (i.e. change the first embedding relative to the second -- an overall change in embedding has no effect). \emph{For each different embedding of H on the first sub-manifold, we get a nontrivially different kink solution which breaks the symmetry in a different way, since at finite $w$ the unbroken symmetry is the intersection of the differently embedded symmetries of the two vacuum sub-manifolds.} All kinks interpolating between the same two sub-manifolds belong into the same \emph{topology class}.

\subsubsection{The asymmetric Kink of the adjoint SO(10) Higgs}
\label{subsubsec:edwards}
As pointed out in the previous Subsection, we cannot use soliton solutions for the Higgs in 3+1 dimensions. However, if we introduce an extra dimension, call it $z$, and let the Higgs vary only along $z$, soliton solutions are allowed if we can somehow place our universe on the resulting domain wall, as a 3+1 dimensional hypersurface embedded in the 5-dimensional \"uber-space. An observer on that hypersurface, perceiving only 3+1 dimensions, sees no topological defects. As we will see, this gives us a dynamic mechanism to generate one or more branes which represent our 3+1 dimensional universe.

This project will use some of the work done by Shin and Volkas in 2002 \cite{edwards}, which I will summarise here. The Higgs field $\Phi$ is in the adjoint representation of SO(10), i.e. a $10 \times 10$ asymmetric matrix with real entries. 
\begin{equation*}
\Phi = (\phi_{ij}) \qquad \phi_{ij} = -\phi_{ji} \in \mathbb{R} \qquad i,j = 1, \ldots,10
\end{equation*}
We restrict our attention to general SO(10) invariant quartic potentials of the form
\begin{equation} \label{eq:higgspot}
V = \sfrac{1}{2} \mu^2 \textrm{Tr}( \Phi^2) + \sfrac{1}{4} \lambda_1 \textrm{Tr}( \Phi^2)^2 + \sfrac{1}{4} \lambda_2 \textrm{Tr}( \Phi)^4.
\end{equation}
The conventionally normalised kinetic energy term is
\begin{equation}
T = -\sfrac{1}{4}\textrm{Tr}(\partial_\mu \Phi \partial^\mu \Phi),
\end{equation}
with Minkowski Signature $(+,-,-,-)$, where the factor out front is $\sfrac{1}{4}$ instead of $\sfrac{1}{2}$ since each independent term is counted twice.
The cubic term in the potential vanishes identically due to the asymmetry of $\Phi$. So in addition to the continuous SO(10) symmetry, the potential has an ``accidental" $\mathbb{Z}_2$ symmetry
\begin{equation*}
\Phi \longrightarrow -\Phi
\end{equation*}
\emph{which means we can use the Clash of Symmetries Mechanism}. As shown in \cite{edwards}, without loss of generality we can assume $\Phi$ to be in the standard form
\begin{equation}
\Phi = \textrm{diag}(a_1 \epsilon, a_2 \epsilon,\ldots, a_5 \epsilon)
\end{equation}
where the $a_i$ are real numbers and 
\begin{equation}
\epsilon = i\sigma_2 = \left( \begin{array}{cc} 0&1\\-1&0\end{array} \right).
\end{equation}
In this basis, the global minima of the potential for $\lambda_2 > 0$ are given by (see \cite{li})
\begin{align*}
a_i^2 &= a_{min}^2  \qquad \forall i \\
\textrm{where} \qquad a_{min} &= \sqrt{\mu^2/(10\lambda_1 + \lambda_2)}.
\end{align*}
For the different minima, adopt the obvious notation $a_{min}\textrm{diag}(-\epsilon,\epsilon,\epsilon,\epsilon,\epsilon) \longrightarrow (-,+,+,+,+)$ and so forth. These minima are invariant under global U(5) transformations embedded in SO(10) (for details see Appendix \ref{ap:embeddings}). The U(5) transformations are embedded entry-wise via the mapping $h$, which takes each complex number in the $5 \times 5$ U(5) matrix to the corresponding $2 \times 2$ subblock in the $10\times 10$ SO(10) matrix:
\begin{equation*}
r e^{i\theta} \stackrel{h}{\longrightarrow}
r \left(\begin{array}{cc}
\cos{\theta} & -\sin{\theta}\\
\sin{\theta} & \cos{\theta} \end{array} \right) \qquad r \geq 0
\end{equation*}
This mapping preserves both multiplicative and additive structure. There are no SO(10) transformations which can change the sign of an odd number of the diagonal $\epsilon$'s, and we can hence divide all possible vacua into two distinct classes (i.e. vacuum sub manifolds) which are related by an overall change in sign.
\begin{align*}
\textrm{Class 1:}& \qquad (+,+,+,+,+), \qquad (+,+,+,-,-), \qquad (+,-,-,-,-)\qquad\textrm{and permutations of the }+/-\textrm{ entries}\\
\textrm{Class 2:}& \qquad (-,-,-,-,-), \qquad (-,-,-,+,+), \qquad (-,+,+,+,+)\qquad\textrm{and permutations of the }+/-\textrm{ entries}
\end{align*}
If the $\Phi(-\infty)$ and $\Phi(\infty)$ vacua belong to different classes, the Higgs will take the form of a \emph{kink} solution with a domain wall at $z = 0$. Let 
\begin{equation}
\Phi(-\infty) = -a_{min}\textrm{diag}(\epsilon,\epsilon,\epsilon,\epsilon,\epsilon).
\end{equation}
The three nontrivially different choices for $\Phi(\infty)$ are
\begin{equation} 
\label{eq:differentkinks}
\Phi(\infty) = 
\begin{cases}
\Phi_{min}^{(5)} = a_{min}\textrm{diag}(\epsilon,\epsilon,\epsilon,\epsilon,\epsilon) & \text{symmetric kink} \\
\Phi_{min}^{(3,2)} = a_{min}\textrm{diag}(\epsilon,\epsilon,\epsilon,-\epsilon,-\epsilon) & \text{asymmetric kink} \\
\Phi_{min}^{(4,1)} = a_{min}\textrm{diag}(\epsilon,-\epsilon,-\epsilon,-\epsilon,-\epsilon) & \text{superasymmetric kink}
\end{cases}
\end{equation}
The symmetry at $z = \pm \infty$ is always a U(5), but the embedding at $z = - \infty$ is different from the embedding at $z = + \infty$. Hence for finite $z$, i.e. in the bulk, the symmetry is the intersection of the two $U(5)$'s. The superscripts in Equation \ref{eq:differentkinks} indicate the bulk symmetry, which for the symmetric, asymmetric and superasymmetric kinks is $U(5)$, $U(3) \otimes U(2)$ and $U(4) \otimes U(1)$ respectively. As we can see, using the Clash of Symmetries Mechanism the SO(10) has been broken down to pleasingly small subgroups using only one Higgs, far more efficiently than with conventional mechanisms. 

The bulk symmetry for the asymmetric kink is $G_{SM} \otimes U(1)$, which is very interesting for model building and justifies making it the focus of our project. The additional $U(1)$ symmetry would represent another ``force" mediated by a Z-boson-like object, which would have to be given a mass at the TeV scale somehow. There is plenty of established theory available for dealing with this extra $U(1)$, and since we are constructing a proof-of-concept toy model we will ignore its ramifications and concentrate on the $G_{SM}$ symmetry.

The embedding of $G_{SM} \otimes U(1)$ is realised by first embedding $U(5)$ in SO(10) by restricting $O \in SO(10)$ to belong to the image of $h$, i.e.
\begin{equation} \label{eq:OtoSU(5)}
O = h(U)
\end{equation}
for some $U \in U(5)$. $U(5) = SU(5) \otimes U(1)'$, so writing out $U(1)'$ charge explicitly we say that $U \in SU(5)$ in (\ref{eq:OtoSU(5)}), and that each transformation carries a saperate $U(1)'$ charge of 2 (chosen to agree with normalisation in \cite{slansky}). We then embed $SU(3) \otimes SU(2) \otimes U(1)''$ in $SU(5)$ by restricting $U \in SU(5)$ to have the form
\begin{equation} \label{eq:UtoGSM}
U = \left(\begin{array}{cc}
U_3 & 0\\
0 & U_2 \end{array} \right).
\end{equation}
The $U_3$ and $U_2$ sub-matrices are $SU(3)$ and $SU(2)$ transformation matrices which carry a $U(1)''$ charge of $-2$ and $3$ respectively. If we then apply the mapping $h$ to this $U$ with $U(1)'$ charge -2, we have the embedding of $SU(3) \otimes SU(2) \otimes U(1)'' \otimes U(1)' = G_{SM} \otimes U(1)$ respected by the asymmetric kink (for more technical detail see Appendix \ref{ap:embeddings}).

The Euler-Lagrange Equations applied to $\mathscr{L} = T - V$ for the different components of the kinks yield
\begin{align}
a_1 = a_2 = a_3 = a_4 = a_5 = f \qquad \phantom{,g} & \qquad \textrm{symmetric kink} \nonumber\\
a_1 = a_2 = a_3 = f,  \qquad a_4 = a_5 = g & \qquad \textrm{asymmetric kink}\nonumber\\
a_1 = f, \qquad a_2 = a_3 = a_4 = a_5 = g & \qquad \textrm{superasymmetric kink}
\end{align}
where $f$ and $g$ are certain odd and even functions of $z$ respectively. Restricting ourselves to the special case $\lambda_1 = 0$, we can get a simple analytical solution
\begin{equation}
f(z) = a_{min} \tanh{(\mu z)}, \qquad g(z) = -a_{min}
\end{equation}
\emph{This is the solution for the asymmetric kink we will use in this project.} Note that since $f(0) = 0$, the symmetry at $z = 0$ is enhanced to $O(6) \otimes U(2)$, which is isomorphic to $SU(4) \otimes SU(2) \otimes U(1)$. This means that there is a Pati-Salam like group \cite{patisalam} at the centre of the domain wall. However, any particles which might be localised at z = 0 would experience slight leakage into the bulk, and since the domain wall is a set of measure zero the left-over symmetry of interest is in fact $G_{SM} \otimes U(1)$.

\emph{The plan-of-attack for this project can now be coherently outlined: We will insert fermions in different SO(10) representations into the Lagrangian and Yukawa-couple them to the Higgs in the asymmetric kink configuration. This will dynamically generate branes with confined chiral fermions via the confinement mechanism discussed in \ref{subsubsec:dynamicconfinementmechanisms}. We hope to find SO(10) fermion representation for which the Standard Model particles are confined with correct 4-dimensional chirality.}

One final thing to point out: For this analytic solution, the asymmetric kink is not stable since the superasymmetric kink has lower energy density. However, for the general case of $\lambda_1 \ne 0$ it is possible that the asymmetric kink emerges as the energetically favoured solution, and it is in this hope that we proceed with the simple analytic solution as a ``proof-of-concept" toy-model to see whether this approach has any merit.

\section{Developing the Group Theoretical Machinery}
\label{sec:theory}
The basic premise of this project is fairly straightforward. We want to develop a toy model of an SO(10) Grand Unified Theory, utilising the Clash of Symmetries mechanism based on the asymmetric kink solution for the Higgs in the 45 of SO(10), which breaks the symmetry down from SO(10) to $G_{SM} \otimes U(1)$  (see \ref{subsubsec:edwards}). Working within the scope of a proof-of-concept toy model we only insert fermions, no gauge fields. We will insert fermion kinetic terms and fermion-Higgs Yukawa-coupling terms into the Lagrangian, and via the fermion confinement mechanism outlined in \ref{subsubsec:dynamicconfinementmechanisms} we hope to confine the Standard Model Fermions. If successful, this theory could be fleshed out to a serious GUT candidate by embedding it in a warped background metric \'a la RS2 \cite{rs2} to include the graviton, and down the line one could attempt to confine gauge fields somehow. 

We will try out different SO(10) representations for the fermions, which we choose based on the criteria that upon restricting the symmetry to the appropriately embedded $G_{SM} \otimes U(1)$, their breakdown representations include the correct representations for the SM particles. We also need to be able to construct a Yukawa-fermion coupling term which is a singlet under SO(10). The representations tested in this project are the $120, 16, 126$ and $144$. An important resource we will use throughout the project is the Group Theory review by Slansky \cite{slansky}, which contains valuable tables on tensor products and breakdowns for the representations of many Lie Groups.

The difficulty lies in the nitty-gritty applied group theory. We have to get our hands dirty and work on the level of individual entries in the representations, to see how exactly the components of the breakdown representations are constructed out of the components of the broken-down SO(10) representation. We then solve the Dirac Equation for all the entries of the SO(10) fermion representation, and from that we can see how the components of the breakdown representations behave in their interaction with the asymmetric Higgs kink. 

Doing such involved group theory calculations with sometimes very large representations at the level of individual entries is not usually done, and searching the literature for the required methods and techniques yielded little valuable information that goes beyond a basic Lie Group Theory course. So apart from any physical results obtained, the work presented here is valuable to anyone attempting this kind of model building in the future.

In this Section we will outline how to perform the required group theoretical breakdown calculations for tensor and spinor representations of SO(10). The actual calculations are then presented in the next section. For all the necessary details on notation and Group Theory in general refer to Appendix \ref{ap:gt}.

\subsection{Breaking down Representations -- General Strategy}
\label{subsec:breakdownstrategy}
Say we have some group $G$ and want to see how one of its representations, call it $R$, breaks down under restriction of $G$ to a subgroup $H$. This means that we want to know how each of the components of $R$ transforms under an $H$-transformation \emph{embedded in $G$} in a specific way. The focus here is on the \emph{transformation properties} of the components.

The strategy for calculating that breakdown is as follows
\begin{compactenum}
\item Determine the smallest representations of $G$ which, via direct product operations, can be used to construct $R$. Call these representations $Y_i$, i.e.
\begin{equation} \label{eq:Rdp}
R = Y_1 \otimes Y_2 \otimes \ldots
\end{equation}
where there are additional conditions imposed on the direct product to match up with the specific symmetry, trace or whatever properties of $R$. 
\item Determine exactly how each of the $Y_i$'s breaks down under restriction of $G$ to $H$ by acting on them with an embedded $H$ transformation and keeping track of all the entries. From these entries one can through linear combinations form groups which transform amongst themselves like representations of H. This will give us a result of the form
\begin{align} \label{eq:Ybd}
G &\supset H \nonumber \\
Y_i &\longrightarrow X(1)_i \oplus X(2)_i \oplus \ldots
\end{align}
for each of the $Y_i$'s, where the $X$'s are representations of H, and we know how the components of the $X$'s are obtained from the components of the $Y$'s. 
\item We then calculate the following direct product of the $X$'s
\begin{align} \label{eq:calcRbd}
G &\supset H  \nonumber \\
R = Y_1 \otimes Y_2 \otimes \ldots &\longrightarrow [X(1)_1 \oplus X(2)_1 \oplus \ldots] \otimes [X(1)_2 \oplus X(2)_2 \oplus \ldots] \otimes \ldots,
\end{align}
where we impose the same conditions on the direct products of the $X$'s as we do on the direct product of the $Y$'s (in order to make it conform with the specific characteristics of $R$). The result of that direct product calculation is then the breakdown of $R$:
\begin{align*}
G &\supset H\\
R &\longrightarrow W_1 \oplus W_2 \oplus \ldots
\end{align*}
where each of the $W$'s is some direct product of $X$'s. 
When performing the direct product in Equation \ref{eq:calcRbd}, we keep track how the components of the $X$'s combine to give the components of the $W$'s. Since we know how the components of the $X$'s are derived from the components of the $Y$'s, and since we know how the components of the $Y$'s combine to transform like corresponding components of the $R$, we know how the components of the $R$'s combine to make up the components of the $W$'s. 
\end{compactenum}
We will now apply this strategy to tensor and spinor representations of SO(N) and tensor representations of SU(N)

\subsection{Breaking down a large SO(N) Tensor Representation}
We will now show how the above procedure is applied to SO(N) tensor representations. We will not repeat all the relations, they are implied by using the same notation.
\begin{compactenum}
\item The smallest tensor representation of SO(N) is $T^i$, the $N$ (so all the $Y$'s are $N$'s). There is no trick to figuring out how $N$ breaks down under restriction of $G = SO(N)$ to $H$. One simply applies the embedded transformation and sees how the components break up. \emph{If for some reason this is nontrivial, the fool-proof method outlined in \ref{subsec:howtobreakdownspinor} also works for tensor representations.}
\item Let $T^{i_1,i_2,\ldots,i_n}$ be the tensor representation $R$, a rank $n$ traceless tensor with certain symmetry properties across its indices. $R$ can be obtained via
\begin{equation*}
R = N \otimes N \otimes N \otimes \ldots
\end{equation*}
where the direct product is performed $n$ times and certain (anti)symmetrising and trace-removing operations are implied. Now 
\begin{equation*} 
T^{i_1,i_2,\ldots,i_n} \thicksim T^{i_1} T^{i_2} \ldots T^{i_n},
\end{equation*}
so for the sake of deriving transformation properties \emph{only}\footnote{To phrase it more precisely: we say that the additional terms ``actually'' present on the LHS of Equation \ref{eq:TtoN} (like (anti)symmetrising and trace-removing terms) are always implied but never actually written, and if we are consistent the calculation works the same way as it would were we to include all the terms. Equation \ref{eq:TtoN} is really just an example of efficient notation to simplify our calculations.}, we can write 
\begin{equation} \label{eq:TtoN}
T^{i_1,i_2,\ldots,i_n} = T^{i_1} T^{i_2} \ldots T^{i_n}.
\end{equation}
We now know how each of the components of the $N$'s combine to form objects which transform like corresponding components of the $R$. 
\item We then perform the tensor product amongst the breakdown reps of the $N$'s. The entries of the resulting $H$-representations will be given by linear combinations of products of $N$-entries of the form $T^{i_1} T^{i_2} \ldots T^{i_n}$, which, using Equation \ref{eq:TtoN}, we simply replace by the corresponding $T^{i_1,i_2,\ldots,i_n}$'s. Hence we have expressed the elements of breakdown products of $R$ in terms of the elements of $R$, and we are done.
\end{compactenum}

\subsection{Breaking down SU(N) Tensor Representations}
The procedure is \emph{exactly} analogous to the one for SO(N) tensor reps, except that the general form of $R$ is 
\begin{equation*}
T^{i_1,i_2,\ldots,i_n}_{j_1,j_2,\ldots,j_m},
\end{equation*}
a rank $(n,m)$ tensor, traceless across any combination of one top and one bottom index, with certain symmetry properties across top and bottom indices separately. It is obtained via 
\begin{equation*}
R = N \otimes N \otimes N \otimes \ldots \otimes \overline{N} \otimes \overline{N} \otimes \overline{N} \otimes \ldots,
\end{equation*}
where the $N$ and $\overline{N}$ occur $n$ and $m$ times respectively, and for the sake of transformation properties \emph{only} we can write 
\begin{equation*}
T^{i_1,i_2,\ldots,i_n}_{j_1,j_2,\ldots,j_m} = T^{i_1} T^{i_2} \ldots T^{i_n} T_{j_1} T_{j_2} \ldots T_{j_m}.
\end{equation*}

\subsection{Obtaining the basic Spinor Representation of SO(2n)}
\label{subsec:howtobreakdownspinors}
Spinor representations are a lot more work than tensor representations. Before we can start to break one down we have to first understand its construction. There was some literature available to help understand the \emph{basic} principles behind Clifford Algebras and the construction of the S matrix, but everything beyond that was derived by actually working with the spinor representations.

\subsubsection{Clifford Algebras}
The rotation group SO(2n) can be considered as the group of linear transformations on the group of coordinates $x_1,x_2,\ldots,x_{2n}$ that leaves the bilinear form
\begin{equation*} \label{eq:basicbilinear}
(x^1)^2 + (x^2)^2 + \ldots + (x^{2n})^2
\end{equation*}
invariant. If we want to rewrite this bilinear form as the square of some general linear form
\begin{equation} \label{eq:squaredlinearform}
(\gamma^1 x^1 + \gamma^2 x^2 + \ldots \gamma^{2n} x^{2n})^2,
\end{equation}
we must require that the $\gamma$'s satisfy the anticommutation relation
\begin{equation} \label{eq:gammacommutation}
\{\gamma^i, \gamma^j\} = 2 \delta^{ij}.
\end{equation}
This collection of $\gamma$ objects is the \emph{Clifford Algebra} of SO(2n). One possible matrix representation of the $\gamma$'s is
\begin{align} \label{eq:gammaform}
\gamma^{2k-1} &= 1 \otimes 1 \otimes \ldots \otimes 1 \otimes \sigma_1 \otimes \sigma_3 \otimes \ldots \otimes \sigma_3 \nonumber \\
\gamma^{2k\phantom{-1}} &= \underbrace{1 \otimes 1 \otimes \ldots \otimes 1}_{k-1} \, \otimes \; \sigma_2 \otimes \underbrace{\sigma_3 \otimes \ldots \otimes \sigma_3}_{n-k}
\end{align}
for $k = 1,\ldots,n$ \cite{zee}, where the $\sigma$'s are the usual $2 \times 2$ Pauli Matrices
\begin{equation*} \label{eq:paulis}
\sigma_1 = \left( \begin{array}{cc} 0&1\\1&0 \end{array}\right) \qquad \sigma_2 = \left( \begin{array}{cc} 0&-i\\i&0 \end{array}\right) \qquad \sigma_3 = \left( \begin{array}{cc} 1&0\\0&-1 \end{array}\right) \qquad 1 = \left( \begin{array}{cc} 1&0\\0&1 \end{array}\right) .
\end{equation*}
It is necessary to introduce two more objects before moving on to spinor representations. Firstly, there is a matrix which anticommutes with all the matrices of the Clifford Algebra:
\begin{equation*} \label{eq:gamma5}
\gamma^{\textrm{FIVE}} = (-i)^n \gamma^1 \gamma^2 \ldots \gamma^{2n} =  \underbrace{\sigma_3 \otimes \ldots \otimes \sigma_3}_{n}.
\end{equation*}
Secondly, define
\begin{equation} \label{eq:sigmaijs}
\sigma^{ij} = \sfrac{i}{2} [\gamma^i, \gamma^j].
\end{equation}
It is straightforward to establish that the $\sigma^{ij}$'s obey the same commutation relation (Equation \ref{eq:Jcommreln}) as the SO(2n) generators $J^{ij}$. The $\sigma^{ij}$ matrices therefore form a $2^n \times 2^n$ representation of the $2n \times 2n$ generators of the defining representation of SO(2n). This means we can use them to obtain a representation of SO(2n), but there is a subtlety which will become clear later.

\subsubsection{Spinors}
Consider what happens to Equation \ref{eq:squaredlinearform} if we apply an SO(2n) rotation in the form of the $2n \times 2n$ transformation matrix $O^i_j$:
\begin{equation*} \label{eq:transformedsquaredlinearform}
(\gamma^i x^i)^2 \longrightarrow (\gamma^i O^i_j x^j) = (O^j_i \gamma^j x^i).
\end{equation*}
As we can see, rotating the $x$'s is equivalent to transforming the $\gamma$'s via
\begin{equation} \label{eq:gammatransform}
\gamma^i \longrightarrow {\gamma'}^i = O^j_i \gamma^j
\end{equation}
These new matrices obviously still satisfy the commutation relation \ref{eq:gammacommutation}, which means they are also a representation of the Clifford Algebra and hence related to the original $\gamma$ matrices by a unitary similarity transformation $S(O)$:
\begin{equation*}
{\gamma'}^i = S(O) \gamma^i S^{-1}(O).
\end{equation*}
Omit the $(O)$ from now on, but keep in mind that it is always implied. Note that $S^\dagger = S^{-1}$ (unitary). Rewrite the above equation as
\begin{equation} \label{eq:Scondition}
O^j_i \gamma^i = S \gamma^i S^{-1}
\end{equation}
which is the equation we can use to determine the form of the $S$ matrix for a given $O$ matrix. The S matrices define a representation of the rotation group. It is called the \emph{spinor representation} of SO(2n). Note that this is a \emph{complex} representation of SO(2n).

As pointed out in Appendix \ref{ap:gt}, a representation can be obtained by arranging all the independent components of some object in a vector $V^i$ and writing the transformation as
\begin{equation*}
V^i \rightarrow M^i_a V^a.
\end{equation*}
where the $M$ matrices form a representation of the group. We apply the reverse process here - we have a set of matrices $S$ which form a representation of SO(2n). We can therefore introduce a vector-type object $\Psi$ with $2^{2n}$ components which transforms with the $S$:
\begin{equation} \label{eq:psitransform}
\Psi^i \longrightarrow S^i_j \Psi^j.
\end{equation}
This object $\Psi$ is called the SO(2n) \emph{spinor}.

\subsubsection{Obtaining the Matrix $S$}
The generators of the fundamental representation of SO(2n) are the $J^{ij}$'s (see Appendix \ref{ap:gt}). $J^{ij}$ is the matrix generating a rotation in the $ij$-plane:
\begin{equation*}
J^{ij} = \left( 
\begin{array}{ccccccc}
0 \\
& \ddots \\
& & 0 & & i\\
&&& \ddots \\
&& -i && 0\\
&&&&& \ddots \\
&&&&&& 0
\end{array} \right)
\end{equation*}
where the $i$ and $-i$ (complex $i$) sit in the $(i,j)$ and $(j,i)$ indices respectively\footnote{Confusion between the index $i$ and complex $i$ should never arise since the index $i$ will never appear in an equation as anything other than a labelling index.}. Note that the $ij$-index is a \emph{label} for the entire matrix, not an entry within a matrix. Writing $J^{ij}$ in index form
\begin{equation} \label{eq:Jindexform}
(J^{ij})^a_b = i(\delta_i^a \delta^j_b - \delta_j^a \delta^i_b)
\end{equation}
it is straightforward to determine that\footnote{Note that the matrix-labelling indices $(i,j)$ run from $1$ to $2n$, whereas the matrix-entry indices $(a,b)$ run from $1$ to $2^{2n}$ -- they act on completely different spaces.}
\begin{equation} \label{eq:Jsquaredcubed}
([J^{ij}]^2)^a_b = \delta^a_b(\delta^a_i + \delta^a_j) \qquad \textrm{and} \qquad (J^{ij})^3 = J^{ij} \qquad \textrm{(no summation over $i,j$)}.
\end{equation}
It is now easy to confirm that $J^{ij}$ generates a rotation in the $ij$-plane by computing the Taylor Expansion of $e^{i \theta J^{ij}}$ using Equation \ref{eq:Jsquaredcubed}.
\begin{equation*}
e^{i \theta J^{ij}} = \left( 
\begin{array}{ccccccccccc}
1 \\
& \ddots \\
&& 1\\
&&& \cos{\theta} &&&& -\sin{\theta}\\
&&&& 1\\
&&&&& \ddots \\
&&&&&& 1\\
&&& \sin{\theta} &&&& \cos{\theta} \\
&&&&&&&& 1 \\
&&&&&&&&& \ddots \\
&&&&&&&&&& 1
\end{array} \right)
\end{equation*}
We can therefore write a general infinitesimal transformation as 
\begin{equation} \label{eq:infintesimalO}
O \simeq 1 + \omega^{ij} \epsilon^{ij}
\end{equation}
where we sum over $(i,j)$, the $\omega$'s are (in this case infinitesimal) rotation angles and 
\begin{equation*}
(\epsilon^{ij})^a_b = (\delta_j^a \delta^i_b - \delta_i^a \delta^j_b),
\end{equation*}
or writing it in matrix form:
\begin{equation} \label{eq:epsilonmatrix}
\epsilon^{ij} = i J^{ij} = \left( 
\begin{array}{ccccccc}
0 \\
& \ddots \\
& & 0 & & -1\\
&&& \ddots \\
&& 1 && 0\\
&&&&& \ddots \\
&&&&&& 0
\end{array} \right).
\end{equation}
Note that $\epsilon^{ij} = - \epsilon^{ji}$. 

By assuming an infinitesimal form for $S$
\begin{equation*}
S \simeq 1 + i w^{ij} S^{ij}
\end{equation*}
and substituting this $S$ and the infinitesimal form of $O$ into Equation \ref{eq:Scondition}, we can find, to first order in $\omega$'s, the form of the $S^{ij}$ matrices:
\begin{align*}
S \gamma^a S^{-1} &= O^b_a \gamma^b \\
\Longrightarrow \qquad \gamma^a + i w^{ij}[S^{ij}, \gamma^a] &= \gamma^a + \omega^{ij} (\epsilon^{ij})^a_b \gamma^b \\
\Longrightarrow  \qquad \phantom{\gamma^a + w^{ij}} i[S^{ij}, \gamma^a] &= -(\epsilon^{ij})^a_b \gamma^b = (\delta_i^a \delta^j_b - \delta_j^a \delta^i_b)\gamma^b = (\delta_i^a \gamma^j - \delta_j^a \gamma^i)
\end{align*}
The form $S^{ij}$ must assume in order to satisfy that commutation relation is
\begin{equation*}
S^{ij} = \sfrac{i}{4} [\gamma^i, \gamma^j] = -\sfrac{i}{2} \sigma^{ij}.
\end{equation*}
Therefore
\begin{equation} \label{eq:Ssolution}
O \simeq 1 + \omega^{ij}\epsilon^{ij} \qquad \Longrightarrow \qquad S \simeq 1 - \sfrac{1}{4} \omega^{ij} [\gamma^i, \gamma^j] \qquad  = 1 + \sfrac{i}{2} \omega^{ij}\sigma^{ij}.
\end{equation}
This now leads us to a very interesting observation about spinor representations. For finite rotation angles $\omega^{ij}$, 
\begin{equation*}
S = e^{\sfrac{i}{2}w^{ij} \sigma^{ij}}.
\end{equation*}
\emph{The factor $\sfrac{1}{2}$ means that only a rotation by an angle $4 \pi$ is equal to the identity, not $2 \pi$.} This is a defining characteristic of Spinors.

\subsubsection{Two distinct basic irreducible Spinors for SO(2n)}
Since $S$ is made up of $\gamma$'s, it anticommutes with $\gamma^{\textrm{FIVE}}$. Therefore we can use $\gamma^{\textrm{FIVE}}$ to project out two components of the $2^n$-component spinor $\Psi$:
\begin{equation} \label{eq:gamma5projections}
\Psi_A = \sfrac{1}{2}(1  - \gamma^{\textrm{FIVE}})\Psi \qquad \qquad \Psi_B = \sfrac{1}{2}(1  + \gamma^{\textrm{FIVE}})\Psi
\end{equation}
where the $\Psi_{A \textrm{ or } B}$ are eigenstates of $\gamma^{\textrm{FIVE}}$:
\begin{equation} \label{eq:gamma5eigenstates}
\gamma^{\textrm{FIVE}}\Psi_A = -\Psi_A \qquad \qquad \gamma^{\textrm{FIVE}}\Psi_B = \Psi_B
\end{equation}
$\Psi_{A \textrm{ or } B}$ are $2^n$-component vectors acted on by a $2^n \times 2^n$ $S$ matrix, but since half their components are zero or not independent we can write the action of $S$ on $\Psi_{A \textrm{ or } B}$  as
\begin{equation*}
\Psi_A \rightarrow S_A \Psi_A \qquad \qquad \Psi_B \rightarrow S_B \Psi_B
\end{equation*}
where the $\Psi_{A \textrm{ or } B}$ are now $2^{n-1}$-component vectors and the $S_A$ and $S_B$ matrices are $2^{n-1} \times 2^{n-1}$ sub-matrices of $S$. It was found that they are related to each other via complex conjugation and some unitary basis change:
\begin{equation} \label{eq:SASBrelation}
S_A = M_S S_B^* M_S^{-1},
\end{equation}
meaning that $\Psi_A$ and $\Psi_B$ form conjugate representations, labelled by $2^{n-1}$ and $\overline{2^{n-1}}$ (or vice versa).

\subsection{Breaking down the basic Spinor Representation of SO(2n)}
\label{subsec:howtobreakdownspinor}
In this project we only have to break down the most basic SO(10) spinor rep, the $16$ (the $144$ and $126$, which are also spinor reps, are analysed using a different method -- more on that later). Therefore, in working with the strategy from \ref{subsec:breakdownstrategy}, we only have to master step 1. 

Breaking down the basic SO(N) tensor representation $N$ under restriction to subgroup $H$ can often be done by inspection, but since any ``natural" embedding of some $H$-transformation in $O$ is scrambled up in $S$, we have to develop a systematic approach. Note that since we only have to do step 1 of the strategy, we will \emph{not} use the same notation as \ref{subsec:breakdownstrategy}.

This approach relies on already knowing the $H$-representations that the SO(2n)-spinor representation, call it $P$, breaks up into. That can be derived using relatively simple group theoretical techniques widely available in the literature, since it has nothing to do with working at the crunchy entry-level of representations. We used the tables in \cite{slansky}. For simplicity, assume $P$ breaks down into only three $H$-representations -- the procedure trivially generalises for more. 

\emph{This procedure can also be applied to tensor representations \footnote{as well as other groups in general} to tell us how the components of its breakdown representations are constructed out of the tensor's entries, if we already know the tensor's breakdown reps.} Here it is:

We know that
\begin{align*}
SO(2n) &\supset H \\
P &\longrightarrow X_1 \oplus X_2 \oplus X_3
\end{align*}
Arrange the independent elements of $X_{1,2,3}$ in vectors $W_{1,2,3}$. We can then write the transformation of the $X$'s under $H$ as
\begin{equation} \label{eq:Wtransform}
W_{1,2,3} \longrightarrow M_{1,2,3} W_{1,2,3}
\end{equation}
for some matrices $M_{1,2,3}$. Similarly, we can arrange the independent elements of $P$ in a vector $V$ and write its transformation under the \emph{embedded} $H$ transformation as
\begin{equation*}
V \longrightarrow M_P V
\end{equation*}
Now rewrite Equation \ref{eq:Wtransform} as
\begin{equation*}
W \longrightarrow M_W W
\end{equation*}
where $M_W$ is in the block-diagonal form
\begin{equation*}
M_W = \left( \begin{array}{ccc} M_1\\ & M_2\\ &&M_3 \end{array}\right) \qquad \textrm{and} \qquad W = \left( \begin{array}{c} W_1 \\ W_2 \\ W_3 \end{array} \right).
\end{equation*}
Since the components of the $X$'s are constructed from linear combinations of the components of the $P$, the matrices $M_W$ and $M_P$ are related by a similarity transformation $Q$:
\begin{equation*}
M_P = Q M_W Q^{-1}
\end{equation*}
which means that 
\begin{equation*}
W = Q^{-1} V.
\end{equation*}
This gives us the exact way that the components of the $X$'s are made up in terms of the components of the $P$.

Note that in order to construct the matrices $M_W$ and $M_P$, we actually have to apply the transformation and explicitly work out how each of its independent components transforms. This is somewhat nontrivial for spinor representations, since the $S$ matrix is an exponential of matrices. However, we can work with infinitesimal transformations, which break up the representations just like finite ones, and for which the explicit form of $S$ is easily obtained.

\section{Inserting Fermions into our Toy-Model}
\label{sec:fermions}
Having developed all the necessary group theoretical machinery, we will now apply it to derive the confinement behaviour of various SO(10) representations for the fermions. We will calculate how each of these SO(10) representations break down under the restriction of SO(10) to $SU(3) \otimes SU(2) \otimes U(1)'' \otimes U(1)' = G_{SM} \otimes U(1)$, which is the symmetry preserved by the asymmetric kink. The specifics of the embedding are explained in \ref{subsubsec:edwards} and Appendix \ref{ap:embeddings}. 

After obtaining the breakdown, we will determine all the possible fermion-Higgs Yukawa coupling terms, insert them into the Lagrangian along with the fermion kinetic term, and solve the resulting Dirac Equation for the fermions to find ``mass eigenstates", i.e. Higgs-coupling eigenstates. We can then determine how each of the breakdown representations couples to the Higgs, and which fermions are confined and where via the confinement mechanism presented in \ref{subsubsec:dynamicconfinementmechanisms}. By identifying some of the appropriate breakdown representations with the first generation of our Standard Model fermions we can see whether our toy model for one fermion generation localises the SM fermions in the right way.

\subsection{The 120}
\label{subsec:120}
The 120 of SO(10) is a rank-3 antisymmetric tensor $T^{ijk}$. The details of the breakdown calculation are straightforward but tedious, and are presented in Appendix \ref{ap:calc120}. Here we will simply state the result:
\begin{equation} \label{eq:120breakdown} 
\begin{array}{l}
SO(10) \supset SU(3) \otimes SU(2) \otimes U(1)'' \otimes U(1)' \\ \\
\phantom{SO}120 \longrightarrow \\ \\
\begin{array}{rccccccccc}
\Big[ & (1,1)(6)(-6) & \oplus & (1,1)(-6)(6) & \oplus & (1,2)(3)(2) & \oplus & (1,2)(-3)(-2) & \oplus\\
 & \scross{6}{IV} & & \scross{5}{II} & & \scross{1}{I} & & \scross{1}{III}\\ \\
 & (1,2)(3)(2) & \oplus & (1,2)(-3)(-2) & \oplus & (3,1)(-2)(2) & \oplus & (\bar{3},1)(2)(-2) & \oplus\\
 & \scross{2}{I} & & \scross{2}{III} & & \scross{1}{II} & & \scross{1}{IV}\\ \\
 & (3,1)(4)(6) & \oplus & (\bar{3},1)(-4)(-6) & \oplus & (3,1)(-2)(2) & \oplus & (\bar{3},1)(2)(-2) & \oplus\\
 & \scross{3}{II} & & \scross{4}{IV} & & \scross{2}{II} & & \scross{2}{IV}\\ \\
 & (3,1)(-8)(-2) & \oplus & (\bar{3},1)(8)(2) & \oplus & (3,2)(1)(-6) & \oplus & (\bar{3},2)(-1)(6) & \oplus\\
 & \scross{4}{II} & & \scross{3}{IV} & & \scross{6}{III} & & \scross{5}{I}\\ \\ 
 & (3,2)(7)(-2) & \oplus & (\bar{3},2)(-7)(2) & \oplus & (3,3)(-2)(2) & \oplus & (\bar{3},3)(2)(-2) & \oplus\\
 & \scross{6}{I} & & \scross{5}{III} & & \scross{9}{I} & & \scross{10}{III}\\ \\ 
 & (6,1)(2)(-2) & \oplus & (\bar{6},1)(-2)(2) & \oplus & (8,2)(3)(2) & \oplus & (8,2)(-3)(-2) & &\Big] \\
 & \scross{6}{II} & & \scross{5}{IV} & & \scross{10}{II} & & \scross{9}{IV} 
\end{array}
\end{array}
\end{equation}
The ``\{arabic numeral\} $\otimes$ roman numeral" labels below each representation indicate the entry of the list in Appendix \ref{ap:120breakdown} which gives the exact form of the entries of that representation in terms of the $T^{ijk}$ tensor elements of the 120.

We now solve the Dirac Equation. The requirement that the Lagrangian be a hermitian singlet under group action gives a unique form for the fermion-Higgs Yukawa-coupling term, which upon substitution of an explicit configuration for $\Phi$ yields a 5-dimensional Dirac mass term  for the fermions (Majorana-like couplings are not discussed, see \ref{subsec:differenthiggscouplings}). The five-dimensional Lagrangian density for our toy-model (omitting Higgs-only terms) is
\begin{equation}
\mathscr{L} = i \Psi^{abc}\Gamma^K\partial_K\Psi^{abc} - ig_{\scriptscriptstyle H}\bar{\Psi}^{abc}\Phi^{ad}\Psi^{dbc}
\end{equation} 
where $g_{\scriptscriptstyle H}$ is a real coupling constant, (a,b,c,d) are SO(10) indices, $K$ is a five-dimensional Lorenz index, $\Gamma^\mu = \tilde{\gamma}^\mu$ and $\Gamma^5 = -i\tilde{\gamma}^5$, and $\Psi^{ijk}$ is the fermion in the 120 of SO(10). The $\tilde\gamma$'s are the usual 4-dimensional $\gamma$-matrices, where we used the tilde to distinguish them from the SO(10) Clifford Algebra. The Dirac Equation for this Lagrangian density (taking into account the antisymmetric property of the $SO(10)$ 120 and 45) is
\begin{equation} \label{eq:120dirac}
0 = i \Gamma^K\partial_K\Psi^{abc} - \frac{ig_{\scriptscriptstyle H}}{3}(\Phi^{ad}\Psi^{dbc} +\Phi^{bd}\Psi^{adc} +\Phi^{cd}\Psi^{abd})
\end{equation}
where we sum over d for any fixed choice of (a,b,c).

We use Mathematica 4.1 to solve the Dirac Equation with the analytic asymmetric kink solution for the Higgs $\Phi$. Firstly, the independent tensor elements of $\Psi^{ijk}$ are arranged in a definite order. Then we substitute the Higgs solution into (\ref{eq:120dirac}) and obtain the coupling terms for each independent $\Psi^{ijk}$. We then construct a $120 \times 120$ mass matrix $M$ to write (\ref{eq:120dirac}) in the following form:
\begin{equation*}
0 = i \Gamma^K\partial_K\ 
\left(\begin{array}{c} \Psi^{1,2,3} \\ \Psi^{1,2,4}\\ \vdots \\ \Psi^{8,9,10} \end{array} \right) +
\left( \begin{array}{ccccc} \phantom{\Psi^{8,9}}\\ & \phantom{\Psi^{8,9}} \\ & & M \\ & & & \phantom{\Psi^{8,9}} \\ & & & & \phantom{\Psi^{8,9}} \end{array} \right)
\left(\begin{array}{c} \Psi^{1,2,3} \\ \Psi^{1,2,4}\\ \vdots \\ \Psi^{8,9,10} \end{array} \right).
\end{equation*}
Upon diagonalising this matrix the mass eigenstate basis is obtained, in which the above equation is written as
\begin{equation}\label{eq:120Cs}
0 = i \Gamma^K\partial_K\ 
\left(\begin{array}{c} \Psi_1 \\ \Psi_2\\ \vdots \\ \Psi_{120} \end{array} \right) + 
\left(\begin{array}{cccc} C_1 \\ & C_2 \\ & & \ddots \\ &&& C_{120} \end{array} \right)
\left(\begin{array}{c} \Psi_1 \\ \Psi_2\\ \vdots \\ \Psi_{120} \end{array} \right)
\end{equation}
where the $C_i$ is the equivalent of the scalar $g \phi$ in \ref{subsubsec:dynamicconfinementmechanisms} for each of the 120 mass eigenstates $\Psi_i$. Since each $C_i$ is a linear combination of the entries of the analytic asymmetric Higgs kink,
\begin{equation*}
f(z) = a_{min} \tanh{(\mu z)}, \qquad g(z) = -a_{min},
\end{equation*}
we are now in a position to analyse the confinement characteristics of each of the mass eigenstates. If $C_i(z_0) = 0$ for some $z_0 \neq \pm \infty$, that mass eigenstate will be confined at $z_0$ via the confinement mechanism outlined in \ref{subsubsec:dynamicconfinementmechanisms}. If the slope of $C_i(z)$ at $z_0$ is positive/negative, the confined zero-mode is right-handed/left-handed.

Using the complete version of Equation \ref{eq:120breakdown} in Appendix \ref{ap:calc120}, we evaluate every element of each of the $SU(3) \otimes SU(2) \otimes U(1)'' \otimes U(1)'$ representations in \ref{eq:120breakdown} and note the $C_i$ for each element. All the elements of a representation have the same $C_i$ as expected, providing a consistency check for our calculation. We now rewrite Equation \ref{eq:120breakdown}, writing under each representation its $C_i$ in terms of $f(z)$ and $g(z)$ (and omitting the common factor $g_{\scriptscriptstyle H}$).

\begin{equation} \label{eq:120breakdownwithcouplings} 
\begin{array}{l}
SO(10) \supset SU(3) \otimes SU(2) \otimes U(1)'' \otimes U(1) \\ \\
\phantom{SO}120 \longrightarrow \\ \\
\begin{array}{rccccccccc}
\Big[ & (1,1)(6)(-6) & \oplus & (1,1)(-6)(6) & \oplus & (1,2)(3)(2) & \oplus & (1,2)(-3)(-2) & \oplus\\
 & f & & -f & & \sfrac{1}{3}g & & -\sfrac{1}{3}g\\ \\
 & (1,2)(3)(2) & \oplus & (1,2)(-3)(-2) & \oplus & (3,1)(-2)(2) & \oplus & (\bar{3},1)(2)(-2) & \oplus\\
 & \sfrac{1}{3}g & & -\sfrac{1}{3}g & & -\sfrac{1}{3}f & & \sfrac{1}{3}f\\ \\
 & (3,1)(4)(6) & \oplus & (\bar{3},1)(-4)(-6) & \oplus & (3,1)(-2)(2) & \oplus & (\bar{3},1)(2)(-2) & \oplus\\
 & \sfrac{2}{3}g - \sfrac{1}{3}f & & -\sfrac{2}{3}g + \sfrac{1}{3}f & & -\sfrac{1}{3}f & & \sfrac{1}{3}f\\ \\
 & (3,1)(-8)(-2) & \oplus & (\bar{3},1)(8)(2) & \oplus & (3,2)(1)(-6) & \oplus & (\bar{3},2)(-1)(6) & \oplus\\
 & -\sfrac{2}{3}g - \sfrac{1}{3}f & & \sfrac{2}{3}g + \sfrac{1}{3}f & & -\sfrac{1}{3}g + \sfrac{2}{3}f & & \sfrac{1}{3}g - \sfrac{2}{3}f\\ \\ 
 & (3,2)(7)(-2) & \oplus & (\bar{3},2)(-7)(2) & \oplus & (3,3)(-2)(2) & \oplus & (\bar{3},3)(2)(-2) & \oplus\\
 & \sfrac{1}{3}g + \sfrac{2}{3}f & & -\sfrac{1}{3}g - \sfrac{2}{3}f & & -\sfrac{1}{3}f & & \sfrac{1}{3}f\\ \\ 
 & (6,1)(2)(-2) & \oplus & (\bar{6},1)(-2)(2) & \oplus & (8,2)(3)(2) & \oplus & (8,2)(-3)(-2) & &\Big] \\
 & -\sfrac{1}{3}f & & \sfrac{1}{3}f & & \sfrac{1}{3}g & & -\sfrac{1}{3}g 
\end{array}
\end{array}
\end{equation}
\emph{This is a general result, independent of the form of the solutions for $f(z)$ and $g(z)$, for the asymmetric Higgs kink.} For the analytic solution, those components for which the $C_i$ includes a $\pm f$ have a right/left-handed confined zero mode, \emph{if the factor in front of the $g$ is not larger than the factor in front of the $f$} -- otherwise the function never goes through zero since $|\tanh{z}| \leq 1$. Some of those representations could be confined if we add a bare mass however (i.e. insert a $m\overline\Psi \Psi$ term in the Lagrangian), which would simply add the same constant to all the $C_i$'s.

It is interesting to note that the $SU(4) \otimes SU(2) \otimes U(1)'''' \otimes U(1)'''$ symmetry displayed at $z = 0$ for the analytic Higgs solution\footnote{In general, at $z = z_0$ for whatever value $z_0$ at which $f(z)$ goes through zero.} is completely disrespected. From \cite{slansky} we get the following branching rules (i.e. breakdown patterns):

\begin{align}
SO(10) &\supset SU(4) \otimes SU(2) \otimes U(1)'''' \nonumber \\
120 & \longrightarrow (1,2)(\sfrac{1}{2}) \oplus (1,2)(-\sfrac{1}{2}) \oplus (10,1)(0) \oplus (\overline{10},1)(0) \oplus (6,3)(0) \nonumber \\
& \phantom{\longrightarrow} \oplus (6,1)(-1) \oplus (6,1)(0) \oplus (6,1)(1) \oplus (15,2)(\sfrac{1}{2}) \oplus (15,2)(-\sfrac{1}{2}) \label{eq:SO(10)SU(4)breakdown} \\ \nonumber \\
SU(4) &\supset SU(3) \otimes U(1)''' \nonumber \\
10 &\longrightarrow 1(2) \oplus 3(\sfrac{2}{3}) \oplus 6(-\sfrac{2}{3}) \nonumber \\
6 &\longrightarrow 3(\sfrac{2}{3}) \oplus \bar{3}(-\sfrac{2}{3}) \nonumber \\
15 &\longrightarrow 1(0) \oplus 8(0) \oplus 3(-\sfrac{4}{3}) \oplus \bar{3}(\sfrac{4}{3})  \label{eq:SU(4)SU(3)breakdown}
\end{align}
For example, we can see that the $(10,1)(0)$ from Equation \ref{eq:SO(10)SU(4)breakdown}, under restriction of $SU(4) \otimes SU(2) \otimes U(1)'''$  to $SU(3) \otimes SU(2) \otimes U(1)'''' \otimes U(1)'''$, breaks down as follows:
\begin{equation}
(10,1)(0) \longrightarrow (1,1)(0)(2) \oplus (3,1)(0)(\sfrac{2}{3}) \oplus (6,1)(0)(-\sfrac{2}{3})
\end{equation}
It is easy to match up these $SU(3) \otimes SU(2) \otimes U(1)'''' \otimes U(1)'''$ representations with $SU(3) \otimes SU(2) \otimes U(1)'' \otimes U(1)'$ representations from Equation \ref{eq:120breakdownwithcouplings}. The $U(1)'' \otimes U(1)'$ quantum numbers $x_1, x_2$ are related to the $U(1)'''' \otimes U(1)'''$ quantum numbers $z_1, z_2$ by a basis change. We can immediately identify the $(6,1)$'s with each other, and there are only two choices to match up the $(1,1)$'s. This gives us two completely consistent possibilities for the basis change required to go from $U(1)'' \otimes U(1)'$ to $U(1)'''' \otimes U(1)'''$ 
\begin{equation*}
z_1 = \pm \sfrac{1}{2} (x_1 + x_2) \qquad \qquad z_2 = \sfrac{1}{15}(-2 x_1 + 3 x_2).
\end{equation*}
For \emph{either} of these choices, it is apparent that the asymmetric kink does not respect the $SU(4) \otimes \ldots$ symmetry when it comes to fermion couplings. All the $(6,1)$, $(1,1)$ and $(3,2)$'s couple differently (\ref{eq:120breakdownwithcouplings}), which means that the $SU(4) \otimes \ldots$ representations are completely ``ripped apart" upon restriction to $SU(3) \otimes \ldots$. It makes sense, since the region where the $SU(4) \otimes \ldots$ symmetry is respected is a set of measure zero, and the fermions are confined over a finite region. This confirms our initial prediction that the $SU(3) \otimes \ldots$ symmetry is the important one.

We now match up the SM particles in $SU(3)_c \otimes SU(2)_W \otimes U(1)_Y$ representations  (\ref{eq:smreps}) with the representations in (\ref{eq:120breakdownwithcouplings}). This is straightforward: All the $(1,2)$'s in (\ref{eq:120breakdownwithcouplings}) have the same $U(1)'' \otimes U(1)'$ quantum numbers, and the $(1,1)$ with the correct chirality has to be the right-handed electron. (\ref{eq:120breakdownwithcouplings}) does not include the right-handed neutrino, which is not a big problem since it is ``optional", and there are only two choices for the left-handed quarks in a $(3,2)$. \emph{In order to let the confined zero modes have the same chirality as the SM particles, we have change the kink to the anti-kink, i.e. replace $f,g$ by $-f,-g$.} The correct embedding of the hypercharge is given by
\begin{equation*}
Y = \sfrac{1}{3} x_1
\end{equation*}
and we can make the identification
\begin{equation*}
f_{L} \leftrightarrow (1,2)(3)(2) \qquad   e^c_{R} \leftrightarrow (1,1)(6)(-6)  \qquad 
Q_{L}  \leftrightarrow  (3,2)(1)(-6) \qquad  u^c_{R} \leftrightarrow (\bar{3},1)(2)(-2)  \qquad d^c_{R} \leftrightarrow (\bar{3},1)(-4)(6)
\end{equation*}

In the table below, the breakdown components are listed \emph{for the anti-kink} grouped by confinement location along the 5th axis $z$, indicated by the form of the $C_i$ (with uninteresting terms omitted), with the subscript indicating which 4-D chiral component is confined at the point where the $C_i$ goes through zero, if applicable. 

\begin{equation}
\begin{array}{cc|ccccc}
C_i: &\textrm{const} & 2+\tanh{} & 1 +2\tanh{} & \tanh{} & 1-2\tanh{} & 2-\tanh{} \\
\hline 
&\\
&(1,2)(3)(2) & (3,1)(-8)(-2)_R & (3,2)(7)(-2)_L & (1,1)(6)(-6)_L & (3,2)(1)(-6)_L & (3,1)(4)(6)_R\\
&(1,2)(3)(2) & & & (3,1)(-2)(2)_R & &\\
&(8,2)(3)(2) & & & (3,1)(-2)(2)_R & &\\
&& & & (3,2)(-2)(2)_R & & \\
&& & & (6,1)(2)(-2)_L & & \\
&\\
\hline 
&\\
&(1,2)(-3)(-2) & (\bar{3},1)(8)(2)_L & (\bar{3},2)(-7)(2)_R & (1,1)(-6)(6)_R & (\bar{3},2)(-1)(6)_R & (\bar{3},1)(-4)(-6)_L\\
&(1,2)(-3)(-2) & & & (\bar{3},1)(2)(-2)_L & &\\
&(8,2)(-3)(-2) & & & (\bar{3},1)(2)(-2)_L & &\\
&& & & (\bar{3},2)(2)(-2)_L & & \\
&& & & (\bar{6},1)(-2)(2)_R & & \\
&\\
\end{array}
\end{equation}

What we have here are in fact different \emph{branes} at different points along the extra dimension, where different fermions are localised. This is reminiscent of the \emph{split-fermion model}, in which fermions are confined to different branes, which exponentially suppresses their wave function overlap and hence provides a possible framework for understanding both fermion mass hierarchy and proton stability (for GUT's where proton stability is an issue) \cite{splitfermion}.

However, from a model building point of view, we can see right away that this fermion representation (with Dirac-like Higgs-fermion Yukawa coupling in the Lagrangian) is no good: All $(1,2)(3)(2)$'s, one of which \emph{has} to represent $f_{iL}$ from the Standard Model (\ref{eq:smreps}), couple \emph{only} to the even function $g$, which means that they are never confined anywhere. One could generalise this calculation by including bare mass terms, but since the $(1,2)$'s will never be confined, there is no point in pursuing this fermion representation any further (at least not with this kind of Higgs coupling).

\subsection{The 16}
\label{subsec:16}
We will use the method outlined in \ref{subsec:howtobreakdownspinor} to break down the 16 of $SO(10)$. Mathematica 4.1 was used for all the symbolic evaluation, and since there is little understanding to be gained from the intermediate calculational steps they will not be given here. Rather, we will outline our strategy in detail and then simply state the relevant results.

We want to break down SO(10) to $SU(3) \otimes SU(2) \otimes U(1)'' \otimes U(1)'$. From \cite{slansky}, we know that 
\begin{align} \label{eq:16goesto}
SO(10) &\supset SU(5) \otimes U(1)' \\
16 &\longrightarrow 1(-5) \oplus \overline{5}(3) \oplus 10(-1),
\end{align}
and it is a simple matter to calculate the breakdown of the SU(5) representations under restriction to $SU(3) \otimes SU(2) \otimes U(1)''$. It simply involves applying the appropriately embedded $SU(3) \otimes SU(2) \otimes U(1)''$ transformation discussed previously and keeping track of the components. The 5 is the row-vector $T^i$ and the $10$ is the $5\times 5$ antisymmetric matrix $T^{ij}$. Their breakdown is as follows:
\begin{equation}
\begin{array}{ccccccc}\label{eq:SU(5)5and10breakdown}
5 & \longrightarrow & (3,1)(-2) & \oplus & (1,2)(3) \\
\scriptstyle {T}^i & & \scriptstyle A^{\rho} = {T}^{\rho} & &\scriptstyle  A^{\nu} = {T}^{\nu + 3}\\
&\\
10 & \longrightarrow & (\bar{3},1)(-4) & \oplus & (3,2)(1) & \oplus & (1,1)(6)\\
\scriptstyle {T}^{ij} & & \scriptstyle B_\sigma = \epsilon_{\sigma a b} T^{ab} & &\scriptstyle B^{\sigma \nu} = T^{\sigma, \nu + 3} & & \scriptstyle B = T^{4,5}
\end{array}
\end{equation}
where below each representation it is indicated how its components are made up from the original SU(5) representation.

The difficult part therefore is to compute the breakdown of the 16 of SO(10) under restriction to $SU(5) \otimes U(1)'$. The first step is to embed an infinitesimal U(5) transformation in an infinitesimal SO(10) transformation. Then we can find the corresponding $S$ matrix which transforms the spinor, giving us an $S$ restricted to an embedded $SU(5)$ transformation.

The following steps easily generalise to U(N) embedded in SO(2n). An infinitesimal U(5) transformation is given by 
\begin{equation*} \label{eq:infinitesimal}
U \simeq 1 + \theta^i (i\lambda^a)
\end{equation*}
where the $\lambda$'s are the $5^2 - 1 = 24$ hermetian generators of SU(5) and the $\theta$'s are the 24 infinitesimal group parameters. The overall U(1) charge is embedded by multiplying each $\theta$ by, say, $Q$ and keeping track of it throughout our calculation. There is one problem with this approach, however: to first order, det$(U) = 1$, which means that the information about the U(1) charge is lost. This means that we are only embedding SU(5) in SO(10), and we have to find out the U(1) charge of the breakdown-representations some other way (we already know it from \cite{slansky}, but we would like to confirm it by finding it ourselves for completeness).

We use the $h$ mapping (shown for a single complex number $z$) to embed the infinitesimal $U$ in an infinitesimal $O \in SO(10)$.
\begin{equation*}
h(z) = \left( \begin{array}{cc} \textrm{Re}\;z & -\textrm{Im}\; z \\ \textrm{Im}\; z & \textrm{Re} \; z \end{array} \right) 
\end{equation*}
so the infinitesimal $SU(5)$ transformation embedded in $O \in SO(10)$ is 
\begin{equation} \label{eq:embeddedinfintesimalU}
O \simeq 1 + \theta^a h(i \lambda^a)
\end{equation}
since $h$ preserves additive and multiplicative structure. We simply have to find $h(i \lambda^a)$ for each SU(5) generator. Let us systematically relabel the 24 SU(5) generators in order to find their SO(10) counterpart. In general, for SU(N), we have $N^2 - 1$ $N\times N$ hermetian generator matrices, which we can group into
\begin{itemize}
\item $N-1$ diagonal generators, call them $\lambda^1, \lambda^2, \ldots, \lambda^{N-1}$. 
\begin{equation*} 
\lambda^i = \left(\begin{array}{cccc} &\\ & 1\\ & &\\ & & & -1 \end{array} \right),
\end{equation*}
where the 1 is in the $(i,i)$ position and the $-1$ is in the $(N,N)$ position. In index notation,
\begin{equation*}
(\lambda^i)^a_b = \delta^a_b(-\delta^a_N + \delta^a_i).
\end{equation*}
\item $\sfrac{1}{2} N(N-1)$ real off-diagonal generators, call them $\lambda^{ij}$ for $j > i$ running from 1 to $N$.
\begin{equation*}
\lambda^{ij} = \left( \begin{array}{ccccc} &\\ &&& 1 & \\ &\\ & 1\\ & \end{array} \right)
\end{equation*}
where the 1's are in the $(i,j)$ and $(j,i)$ positions. In index notation,
\begin{equation*}
(\lambda^{ij})^a_b = \delta^a_i \delta^j_b + \delta^a_j \delta^i_b.
\end{equation*}
\item $\sfrac{1}{2} N(N-1)$ imaginary off-diagonal generators, call them $\widetilde{\lambda}^{ij}$ for $j > i$.
\begin{equation*}
\widetilde{\lambda}^{ij} = \left( \begin{array}{ccccc} &\\ &&& -i & \\ &\\ & i\\ & \end{array} \right)
\end{equation*}
where the $-i$ and $i$ are in the $(i,j)$ and $(j,i)$ positions respectively. In index notation,
\begin{equation*}
(\widetilde{\lambda}^{ij})^a_b = i(- \delta^a_i \delta^j_b + \delta^a_j \delta^i_b).
\end{equation*}
\end{itemize}
where all other entries in the above matrices are zero. In this notation, an infinitesimal $SU(5)$ transformation is
\begin{equation} \label{eq:infintesimalUrelabelled}
U \simeq 1 + \sum_{a=1}^{N-1}\theta^a(i\lambda^a) + \sum_{a<b}^{a,b \leq N} (\theta^{ab} (i\lambda^{ab}) + \widetilde{\theta}^{ab}(i \widetilde{\lambda}^{ab}))
\end{equation}
with appropriately relabelled infinitesimal group parameters. It is then a simple matter to find $h(i\lambda)$ for each kind of generator:
\begin{align*}
h(i\lambda^i) &= \epsilon^{2i-1,2i} - \epsilon^{2N-1,2N} \\
h(i\lambda^{ij}) &= \epsilon^{2i-1,2j} - \epsilon^{2i,2j-1} \\
h(i\widetilde{\lambda}^{ij}) &= -\epsilon^{2i,2j} - \epsilon^{2i-1,2j-1}
\end{align*}
where the $\epsilon$'s are as in (\ref{eq:epsilonmatrix}). Therefore, the SU(5) transformation embedded in SO(10) is
\begin{equation*} 
O \simeq 1 + \sum_{a=1}^{N-1}\theta^a(\epsilon^{2a-1,2a} - \epsilon^{2N-1,2N}) + \sum_{a<b}^{a,b \leq N} (\theta^{ab} (\epsilon^{2a-1,2b} - \epsilon^{2a,2b-1}) + \widetilde{\theta}^{ab}(-\epsilon^{2a,2b} - \epsilon^{2a-1,2b-1}))
\end{equation*}
Finally, we can use Equation \ref{eq:Ssolution} to find the corresponding infinitesimal $S$ matrix:
\begin{equation} \label{eq:SO(10)Smatrix}
S \simeq 1 + \sum_{a=1}^{N-1}\frac{\theta^a}{2}(\gamma^{2a-1}\gamma^{2a} - \gamma^{2N-1}\gamma^{2N}) + \sum_{a<b}^{a,b \leq N} \left[ \frac{\theta^{ab}}{2} (\gamma^{2a-1}\gamma^{2b} - \gamma^{2a}\gamma^{2b-1}) + \frac{\widetilde{\theta}^{ab}}{2}(-\gamma^{2a}\gamma^{2b} - \gamma^{2a-1}\gamma^{2b-1})\right]
\end{equation}
where the $\gamma$ matrices are as per Equation \ref{eq:gammaform} with $n = 5$.

Equipped with this $S$ matrix we can now proceed to transform a 32-component spinor $\Psi$ via 
\begin{equation*}
\Psi^i \longrightarrow S^i_j \Psi^j.
\end{equation*}
This spinor is made up of $\Psi_A$ and $\Psi_B$ as per Equation \ref{eq:gamma5projections}, each of which is a 32-component column vector with 16 empty entries, and the other 16 entries the same as $\Psi$. At this stage we do not know which of these two Spinors is the $16$ and which one is the $\overline{16}$. By acting on $\Psi_{A \textrm{ or } B}$ with S, we find out how its entries transform.

As per Equation \ref{eq:SASBrelation} we then find the relationship between $S_A$ and $S_B$, the $S$-submatrices that act on the different irreducible spinor components. As expected, one obtains one from the other by complex conjugation and by applying a unitary transformation $M_S$ (given in Appendix \ref{ap:16}).

We then transform the SU(5) representations $5$, $\overline{5}$, $10$ and $\overline{10}$ with the infinitesimal $U$ transformation in (\ref{eq:infintesimalUrelabelled}). Knowing that the 16 breaks down into $1 \oplus \overline{5} \oplus 10$, we perform the procedure from \ref{subsec:howtobreakdownspinor}, assuming first that the $\Psi_A$ is the 16, and if that does not work, assume that the $\Psi_B$ is the 16. 

\emph{As it turns out, $\Psi_B$ transforms as the $16$ and the $\Psi_A$ as the $\overline{16}$.} The independent elements of the SU(5) $\overline{5}$, 10 and 1 are embedded in $\Psi_B$ the following way:
\begin{equation} \label{eq:16embedding}
\overline{5}: \left( \begin{array}{c}T_1 \\ T_2 \\ T_4 \\T_4\\T_5\end{array} \right) =
\left( \begin{array}{c}\Psi_{\scriptscriptstyle B}^8 \\ -\Psi_{\scriptscriptstyle B}^{12} \\ \Psi_{\scriptscriptstyle B}^{14} \\-\Psi_{\scriptscriptstyle B}^{15}\\\Psi_{\scriptscriptstyle B}^{16}\end{array} \right) \qquad \qquad
10: \left( \begin{array}{c}T^{12} \\ T^{13}\\ T^{14}\\ T^{15}\\ T^{23}\\ T^{24}\\ T^{25}\\ T^{34}\\ T^{35}\\ T^{45}\end{array} \right) = \left( \begin{array}{c}\Psi_{\scriptscriptstyle B}^{13} \\ \Psi_{\scriptscriptstyle B}^{11}\\ \Psi_{\scriptscriptstyle B}^{10}\\ \Psi_{\scriptscriptstyle B}^9\\ \Psi_{\scriptscriptstyle B}^7\\ \Psi_{\scriptscriptstyle B}^6\\ \Psi_{\scriptscriptstyle B}^5\\ \Psi_{\scriptscriptstyle B}^4\\ \Psi_{\scriptscriptstyle B}^3\\ \Psi_{\scriptscriptstyle B}^2\end{array} \right) \qquad \qquad 1: T = \Psi_{\scriptscriptstyle B}^1.
\end{equation}
Similarly for independent elements of the SU(5) $5$, $\overline{10}$ and 1 are embedded in $\Psi_A$ the following way\footnote{Note that, while (\ref{eq:16barembedding}) was computed using the method from \ref{subsec:howtobreakdownspinor}, we could have also gotten it by complex conjugating all of (\ref{eq:16embedding}) and performing the basis change on the $\Psi$'s using our $M_S$ matrix from (\ref{eq:Sbasischange})}:
\begin{equation} \label{eq:16barembedding}
5: \left( \begin{array}{c}T^1 \\ T^2 \\ T^4 \\T^4\\T^5\end{array} \right) = 
\left( \begin{array}{c}\Psi_{\scriptscriptstyle A}^9 \\ \Psi_{\scriptscriptstyle A}^{5} \\ \Psi_{\scriptscriptstyle A}^{3} \\\Psi_{\scriptscriptstyle A}^{2}\\\Psi_{\scriptscriptstyle A}^{1}\end{array} \right) 
\qquad \qquad
\overline{10}: \left( \begin{array}{c}T_{12} \\ T_{13}\\ T_{14}\\ T_{15}\\ T_{23}\\ T_{24}\\ T_{25}\\ T_{34}\\ T_{35}\\ T_{45}\end{array} \right) = 
\left( \begin{array}{c}\Psi_{\scriptscriptstyle A}^{4} \\ -\Psi_{\scriptscriptstyle A}^{6}\\ \Psi_{\scriptscriptstyle A}^{7}\\ -\Psi_{\scriptscriptstyle A}^8\\ \Psi_{\scriptscriptstyle A}^{10}\\ -\Psi_{\scriptscriptstyle A}^{11}\\ \Psi_{\scriptscriptstyle A}^{12}\\ \Psi_{\scriptscriptstyle A}^{13}\\ -\Psi_{\scriptscriptstyle A}^{14}\\ \Psi_{\scriptscriptstyle A}^{15}\end{array} \right) \qquad \qquad 
1: T = \Psi_{\scriptscriptstyle A}^{16}.
\end{equation}
\emph{This result, combined with (\ref{eq:SU(5)5and10breakdown}), gives us the complete breakdown of the $16$ and the $\overline{16}$ of SO(10) under restriction to $SU(3) \otimes SU(2) \otimes U(1)'' \otimes U(1)'$.} 

We now went on to solve the Dirac Equation for the 16. Once again there is only one choice for the Dirac-like Higgs-fermion Yukawa coupling term. The Lagrangian (with Higgs-only terms omitted) without bare mass terms for the $16$, i.e. $\Psi_B$, is
\begin{equation} \label{eq:16lagrangian}
\mathscr{L} = i \overline{\Psi}_B \Gamma^K \partial_K \Psi_B + i g_{\scriptscriptstyle B} \overline{\Psi}_B \gamma^a \gamma^b \Phi^{ab} \Psi_B
\end{equation}
and we use all the same symbols and notation as for the 120 Lagrangian. The sums running over spinor indices from 1 to 32 have been expressed in matrix/vector notation, where $\Psi_B$ is regarded as a 32-component vector with half its components zero. The Dirac Equation is
\begin{equation} \label{eq:16dirac}
0 = i \Gamma^K \partial_K \Psi_B + i g_{\scriptscriptstyle B} \gamma^a \gamma^b \Phi^{ab} \Psi_B
\end{equation}
Solving this Dirac Equation is a simple matter, since for a $\Phi$ in the standard form, the $\gamma^a \gamma^b \Phi^{ab}$ $32 \times 32$ matrix is in fact already diagonal. The coupling term in (\ref{eq:16dirac}) (i.e. the second term) is, in terms of $f(z)$ and $g(z)$,
\begin{align}
&\textrm{symmetric kink:} \nonumber \\
g_{\scriptscriptstyle B}  \Big[& -10 f \Psi_B^1 , -2 f\Psi_B^2 , -2 f\Psi_B^3 , -2 f\Psi_B^4 ,-2 f \Psi_B^5 , -2 f\Psi_B^6 , -2 f\Psi_B^7 , 6f \Psi_B^8 ,\nonumber \\
&-2 f \Psi_B^9 ,-2 f \Psi_B^{10} , -2 f\Psi_B^{11} , 6f \Psi_B^{12} , -2 f \Psi_B^{13} , 6f\Psi_B^{14} , 6f\Psi_B^{15} , 6f\Psi_B^{16}\Big]^T \label{eq:16symmetriccouplings}\\
&\textrm{asymmetric kink:}  \nonumber \\
g_{\scriptscriptstyle B} \Big[& (4g-6f)\Psi_B^1 , (-4g-6f)\Psi_B^2 ,-2f \Psi_B^3 ,-2f \Psi_B^4 ,-2f \Psi_B^5 , -2f\Psi_B^6 ,(4g+2f) \Psi_B^7 ,(-4g+2f) \Psi_B^8 , \nonumber \\
&-2f \Psi_B^9 , -2f \Psi_B^{10} , (4g+2f) \Psi_B^{11} ,(-4g+2f) \Psi_B^{12} , (4g+2f)\Psi_B^{13} , (-4g+2f)\Psi_B^{14} , 6f\Psi_B^{15} , 6f\Psi_B^{16}\Big]^T \label{eq:16asymmetriccouplings} 
\end{align}
Keep in mind that the symmetric kink preserves the $SU(5) \otimes U(1)''$. \emph{The very interesting thing to note from (\ref{eq:16symmetriccouplings}) is that the couplings to $f$ for each entry are proportional to the $U(1)'$ quantum number of the representation it belongs to.} More on that later.

We write down all the $SU(3) \otimes SU(2) \otimes U(1)'' \otimes U(1)$  breakdown components of the 16 and how they match up with the Standard Model fermions. As before for the 120 breakdown, the hypercharge $Y$ is given by $\sfrac{1}{3} x_1$. Using (\ref{eq:16asymmetriccouplings}), we can write their couplings to the asymmetric kink under each representation:
\begin{align}
SO(10) &\supset  SU(3) \otimes SU(2) \otimes U(1)'' \otimes U(1)' \nonumber \\
16 &\longrightarrow \underbrace{(1,1)(0)(-5)}_{\nu_R^c: \; \; 4g-6f} \oplus \underbrace{(1,1)(6)(-1)}_{e_R^c: \;\;-4g-6f} \oplus \underbrace{(\bar{3},1)(2)(3)}_{d_R^c: \;\;-4g+2f} \oplus \underbrace{(\bar{3},1)(-4)(-1)}_{u_R^c: \;\; 4g+2f} \oplus \underbrace{(3,2)(1)(-1)}_{Q_L:\;\; -2f} \oplus \underbrace{(1,2)(-3)(3)}_{f_L: \;\;6f} \label{eq:16breakdownwithcouplings}
\end{align}
where the couplings given under each representation (e.g. $-2f$ for $Q_L$) are the equivalent of the scalar $g \phi$ in \ref{subsubsec:dynamicconfinementmechanisms} for that fermion. (For the $\overline{16}$, simply conjugate all representations, multiply all the couplings by $-1$ and charge-conjugate all the SM particles.) Once again we have a situation reminiscent of the split-fermion model \cite{splitfermion} as discussed for the 120, with the different confined fermions localised to branes at different locations along the extra dimension.

If we use the analytic solution for the asymmetric kink, it quickly becomes clear that this toy model is not behaving like the standard model should -- all the SM fermions (in the appropriately charge-conjugated basis used here) have to have the same chirality, but according to (\ref{eq:16breakdownwithcouplings}) the confined zero-modes of the $e_R^c$ and $Q_L$ are left-handed whereas all the others are right-handed. Furthermore, the two right-handed quarks are unconfined since, due to the larger factor in front of the $g$, the coupling never goes through zero. 

All hope is not lost  -- we have not yet considered the \emph{most general} Lagrangian for the $16$. We can insert the $\Psi_A$ in charge-conjugate basis, as well as Dirac and Majorana bare mass terms. Our most general Lagrangian is therefore
\begin{align} \label{eq:general16Lagrangian}
\mathscr{L} = & i \overline{\Psi}^c_A \Gamma^K\partial_K \Psi^c_A + i \overline{\Psi}_B \Gamma^K \partial_K \Psi_B + i g_{\scriptscriptstyle A} \overline{\Psi}^c_A \gamma^a \gamma^b \Phi^{ab} \Psi^c_A + i g_{\scriptscriptstyle B} \overline{\Psi}_B \gamma^a \gamma^b \Phi^{ab} \Psi_B  + \nonumber \\
& m_A \overline{\Psi}^c_A \Psi^c_A + m_B \overline\Psi_B \Psi_B + m_{AB}(\overline\Psi_A^c M_S \Psi_B + \overline\Psi_B M_S \Psi^c_A)
\end{align}
where once again we suppress the spinor indices and instead deal with them via matrix/vector notation. Normally we would have to complex conjugate the Higgs coupling term for the $\Psi_A^c$, but since for a Higgs in the standard form that term is completely imaginary, it is still a singlet. Note that for the Majorana Mass term, we have to insert the basis change matrix $M_S$ between the Spinors, in order to match up the proper elements. 

The corresponding Dirac Equation is
\begin{equation} \label{eq:general16dirac}
0 = i \Gamma^K \partial_K \left( \begin{array}{c} \Psi_A^c \\ \Psi_B \end{array} \right) + 
\left(\begin{array}{cc} 
i g_{\scriptscriptstyle A} \gamma^a \gamma^b \Phi^{ab} + m_A  & m_{AB} M_S \\
 m_{AB} M_S & i g_{\scriptscriptstyle B} \gamma^a \gamma^b \Phi^{ab} + m_B 
\end{array} \right) 
\left( \begin{array}{c} \Psi_A^c \\ \Psi_B \end{array} \right)
\end{equation}
and upon diagonalising the matrix in front of the second term we get our mass eigenstates (equivalent to the $C_i$'s from Equation \ref{eq:120Cs} for the 120). As expected, we get two copies of each $SU(3) \otimes SU(2) \otimes U(1)'' \otimes U(1)$ representation produced by the $\Psi_A^c$/$\Psi_B$ breakdown. The entries of each copy are made up of different linear combinations of the elements of $\Psi_A^c$ and $\Psi_B$ corresponding to that representation, so group theoretically everything works out nicely, providing a consistency check of our calculation.

For the general case (i.e. unrestricted values for all coupling constants $g_A$, $g_B$, $m_A$, $m_B$ and $m_{AB}$), Higgs-couplings for each of the eigenstates are quite unwieldy and hard to analyse. They are given in Appendix \ref{ap:16}. However, for the special case $g_A = -g_B = g$, we get couplings which are quite manageable, and on first sight they seem quite promising: for each of the doubled-up representations, one copy is totally unconfined whereas one has the chance of being confined, with not immediately obvious chirality. The couplings for the elements of the doubled-up $SU(3) \otimes SU(2) \otimes U(1)'' \otimes U(1)$ representations are (up to a common factor):
\begin{align} \label{eq:funkyfermioncouplings}
(1,1)(6)(-1): \qquad & (m_A + m_B) \pm \sqrt{4 m_{AB}^2 + (m_A - m_B + 8 + 12 \tanh{(\mu z)})^2 } \nonumber \\
(1,1)(0)(-5): \qquad & (m_A + m_B) \pm \sqrt{4 m_{AB}^2 + (m_A - m_B - 8 + 12 \tanh{(\mu z)})^2 } \nonumber \\
(1,2)(-3)(3): \qquad & (m_A + m_B) \pm \sqrt{4 m_{AB}^2 + (m_A - m_B - 12 \tanh{(\mu z)})^2 } \nonumber \\
(\bar{3},1)(-4)(-1): \qquad & (m_A + m_B) \pm \sqrt{4 m_{AB}^2 + (m_A - m_B - 8 -4 \tanh{(\mu z)})^2 } \nonumber \\
(\bar{3},1)(2)(3): \qquad & (m_A + m_B) \pm \sqrt{4 m_{AB}^2 + (m_A - m_B + 8 -4 \tanh{(\mu z)})^2 } \nonumber \\
(3,2)(1)(-1): \qquad & (m_A + m_B) \pm \sqrt{4 m_{AB}^2 + (m_A - m_B + 4 \tanh{(\mu z)})^2 }
\end{align}
where we have scaled all the bare masses by $1/(g a_{min})$. We can now try and find some values for the mass constants to try and get them all confined with the correct	 chirality (for positive masses, those reps which have the coupling with the $+$ in front of the square root are never confined, so we are always talking about those with the $-$). All couplings are of the form
\begin{equation*}
(m_A + m_B) \pm \sqrt{4 m_{AB}^2 + (m_A - m_B + D_1 + D_2 \tanh{(\mu z)})^2 }
\end{equation*}
for various values of $D_1$ and $D_2$. For this to be zero for \emph{one} of $+$ or $-$, the necessary and sufficient conditions are that
\begin{equation} \label{eq:allconfinedcondition}
(m_A + m_B)^2 \leq 4m_{AB}^2 \qquad \textrm{and} \qquad -1 < \frac{\pm\sqrt{((m_A + m_B)^2 - 4m_{AB}^2)} - m_A + m_B - D_1}{D_2} < 1
\end{equation}
for either $+$ or $-$. If we want one of each of the duplicate representations in (\ref{eq:funkyfermioncouplings}) to be confined, we have to require that (\ref{eq:allconfinedcondition}) is true for all the values $D_1$ and $D_2$ can assume in (\ref{eq:funkyfermioncouplings}) (since if the coupling goes through zero, the fermion is confined there). Working through all the inequalities, we find however that \emph{it is never possible to confine all the fermions in (\ref{eq:funkyfermioncouplings})}.

One could go back and work with the general couplings for each of the representations, and start playing with different values of $g_A$ and $g_B$, but that does not seem worth the effort. Even if one were to find a certain point in parameter space for which all fermions are confined with the correct chirality, that situation would be particular to the analytical solution for the asymmetric Higgs kink. That kink, however, is unstable, since it has a higher energy density than the superasymmetric kink. It would be wiser to find a kink solution for which the asymmetric kink is actually stable, and \emph{then} invest all the effort of trying to find that particular point in the parameter space of the couplings for which all fermions are properly confined. If a proof of concept toy-model requires too much work and fine-tuning to be realistic, maybe one should change some of its premises.


But as already mentioned, the $16$ is not dead yet. For a different, actually stable solution for the asymmetric Higgs kink, things might work out much better.

\subsection{The Big Coincidence}
\label{subsec:coincidence}
We were startled by the fact that the symmetric kink couples to the fermions exactly proportionally to their $U(1)'$ charge. Then we realised: \emph{the $\Phi$ is in the adjoint representation of SO(10), just like a gauge field!} Group theoretically, for an adjoint Higgs, the \emph{Dirac-like} Higgs-fermion Yukawa coupling term is completely equivalent to a gauge field coupling term. Hence, if the Higgs' configuration is such that it mimics a Gauge Field generator that commutes with all the other generators of the unbroken symmetry, it will couple to the charge associated with that generator! $U(1)'$ charge is generated by the generator of U(5) which commutes with all the other generators:
\begin{equation}
X_2 = \left( \begin{array}{ccccc} 2\\ & 2\\ && 2\\ &&&2\\&&&&2 \end{array} \right) \qquad \textrm{with SO(10) equivalent} \qquad
\left( \begin{array}{ccccc} 2\epsilon\\ & 2\epsilon\\ && 2\epsilon\\ &&&2\epsilon\\&&&&2\epsilon \end{array} \right)
\end{equation}
up to an irrelevant normalisation constant. Similarly, when restricting the $SU(5)$ to $SU(3) \otimes SU(2) \otimes U(1)''$, the $U(1)''$ charge is generated by another generator which commutes with all others
\begin{equation}
X_1 = \left( \begin{array}{ccccc} -2\\ & -2\\ && -2\\ &&&3\\&&&&3 \end{array} \right) \qquad \textrm{with SO(10) equivalent} \qquad \left( \begin{array}{ccccc} -2\epsilon\\ & -2\epsilon\\ && -2\epsilon\\ &&&3\epsilon\\&&&&3\epsilon \end{array} \right)
\end{equation}
Which one of these charges does the Higgs couple to? That depends on the configuration. The symmetric kink has the configuration
\begin{equation}
\left( \begin{array}{ccccc} f\epsilon\\ & f\epsilon\\ && f\epsilon\\ &&&f\epsilon\\&&&&f\epsilon \end{array} \right)
\end{equation}
and hence couples to the $U(1)'$ charge, as we observed for the $16$. The asymmetric kink has the configuration
\begin{equation}
\left( \begin{array}{ccccc} f\epsilon\\ & f\epsilon\\ && f\epsilon\\ &&&g\epsilon\\&&&&g\epsilon \end{array} \right)
\end{equation}
and in order to find how the fermions couple to the $f$ and $g$ we have to express the relevant components of the Higgs Kink as linear combinations of the corresponding $U(1)$-generators:
\begin{equation}
g \left( \begin{array}{ccccc} 0\\ & 0\\ && 0\\ &&&10\epsilon\\&&&&10\epsilon \end{array} \right) = g(2X_1 + 2X_2) \qquad \qquad 
f\left( \begin{array}{ccccc} 10\epsilon\\ & 10\epsilon\\ && 10\epsilon\\ &&&0\\&&&&0 \end{array} \right) = f(2X_1 - 3X_2)
\end{equation}
Hence, for each fermion in a representation of $SU(3) \otimes SU(2) \otimes U(1)'' \otimes U(1)'$ interacting with the asymmetric kink, its coupling term, up to a normalisation constant, is given by
\begin{equation} \label{eq:adjointgaugecoupling}
(2x_1+2x_2)g + (2x_1 - 3x_2)f
\end{equation}
where $x_1$, $x_2$ are the $U(1)''$, $U(1)'$ charges, respectively, of the fermion.

This gives us a way to almost instantaneously determine the confinement properties for any fermion representation, \emph{as long as the Higgs is in the adjoint representation and we do not include any bare-mass terms or Majorana-like Higgs-fermion Yukawa coupling terms}. All we need to know is into what representations the original fermion representation breaks down to, without any knowledge of how they are embedded. We can simply use the tables in \cite{slansky}. Furthermore, we can now legitimately use the adjoint Higgs symmetric kink as a U(1)-probe in order to confirm the U(1)-charges of the breakdown representations in (\ref{eq:16goesto}).

One should not interpret this finding as making the group theoretical techniques discussed in this project irrelevant. It might present a shortcut for doing part of a special case of calculations, but we need the heavy group theoretical machinery to perform a general analysis.

\subsection{The 126 and the 144}
\label{subsec:126and144}
Using the ``Big Coincidence", we can very easily perform an analysis of the basic confinement characteristics of the $126$ and the $144$ interacting with the asymmetric kink (without any bare mass terms and only for Dirac-like fermion-Higgs couplings). From \cite{slansky} we know how the $126$ breaks down. Focus on the $SU(2)$ doublets:
\begin{equation*}
126 \longrightarrow (1,2)(-3)(-2) \oplus (1,2)(3)(2) \oplus \ldots
\end{equation*}
Using (\ref{eq:adjointgaugecoupling}), we can see right away that these doublets only couple to $g$, not $f$, meaning they are not confined, making this representation useless from a model-building point of view . Similarly we can quickly look at the $144$. It breaks down into many representations of all shapes and sizes, and while it looks like \emph{maybe} the SM fermions might be confined, we have the same problem with wrong chiralities as with the $16$, and there are a lot of other particles floating around that we have to deal with somehow. Not a very pretty picture.

\subsection{Side Remark: Majorana-like Higgs-Fermion Coupling Terms}
\label{subsec:differenthiggscouplings}
Throughout this project we only considered Dirac-like fermion-Higgs Yukawa coupling terms  $\overline{\Psi} \Phi \Psi$. However, the 120 also allows for a Majorana-like coupling term $\Psi^T C_5 \Phi \Psi + h.c.$, where $C_5$ is the 5-dimensional charge conjugation operator $C_5 = \Gamma^0 \Gamma^2 \Gamma^5$. (The 16 does not, since $16 \otimes 16 \otimes 45$ does not yield a singlet -- similarly for the 126 and 144). \emph{So for the 120, the analysis presented here does not represent the most general case, and that representation is definitely worth exploring further.}

There is a subtle point to make about the discrete $\mathbb{Z}_2$ symmetry $\Phi \rightarrow -\Phi$. We have to associate some kind of discrete action on the $\Psi$ with that symmetry, and the two possible choices are $\Psi \rightarrow i\Psi$ and $\Psi \rightarrow \Gamma^n \Psi$ where we can have $n = 1,2,3,5$ (no sum). The first choice will leave Majorana-like Higgs-fermion coupling terms and Dirac-like bare mass terms invariant, whereas the second choice will leave Dirac-like Higgs-fermion coupling terms and Majorana-like bare mass terms invariant.\footnote{Interestingly, while the first choice leaves the fermion kinetic term invariant, the second choice will induce a parity-like transformation realised by a reflection along the $n$-th spatial axis.}

Having picked our symmetry transformation for the $\Psi$, if we include terms which violate the discrete symmetry then there will be radiative corrections proportional to \emph{odd} powers of $\Phi$ which skew the potential and make the kink configuration for the Higgs unstable. In this case we have to assume this back reaction to be small and the lifetime of the kink to be much larger than the age of the universe.

\section{Conclusion}
\label{sec:conclusion}
The Clash of Symmetries mechanism provides a very efficient and aesthetically pleasing way of breaking a large gauge symmetry down to a much smaller one. In this project we continued the work started by  Shin and Volkas \cite{edwards}, using their solution for the SO(10) Higgs field in the asymmetric kink configuration. We constructed Grand Unified Toy Models with various SO(10) representations for the fermions, which we then dynamically confined to a brane (or different branes) by coupling them to the Higgs field. Although the results were not as promising as we hoped there are several other avenues to explore within the framework constructed here. At any rate, the group theoretical techniques that were developed will be useful to anyone attempting this kind of model building in the future. 

The motivation of this model, in that it combines the efficient symmetry breaking of Higgs Kinks with their ability to dynamically generate branes in a higher-dimensional universe, remains a worthwhile one regardless of the outcome of any follow-up investigations to this project, and it remains possible that these principles may be applied to a different gauge symmetry or kink configuration to eventually construct a true Grand Unified Theory.

\newpage
\appendix

\section{Basic Group Theory Review}
\label{ap:gt}
Most of the material presented in this Appendix is taken from \cite{cheng} and \cite{zee}. Here I will only give a brief overview the group theory required to understand the Standard Model. In Section \ref{sec:theory} I will carry this further to develop the specific techniques required for my research project. I will use Einstein Summation Convention unless otherwise indicated, and 1 denotes the number one or an identity matrix of appropriate size.

\subsection{The Very Basics}
A \emph{group} is a collection of elements $\{g_i\}$ and a group operation $\ast$ such that
\begin{compactenum}[i)]
\item For any $g_i$ in $G$, $g_i \ast g_j$ is also in $G$ (closure)
\item There exists an element $e$ in $G$ such that $e \ast g_i = g_i \ast e = g_i$ (identity)
\item For every element $g_i$ in $G$ there exists an inverse element $g_i^{-1}$ such that $g_i \ast g_i^{-1} = g_i^{-1} \ast g_i = e$ (inverse)
\item $g_i\ast (g_j \ast g_k) = (g_i \ast g_j) \ast g_k$ (associativity)\\
\end{compactenum}
(Omit the $\ast$ from now on and regard it as a "generalised multiplication".) If the group operation is commutative, the group is called \emph{abelian}. A \emph{subgroup} $K$ of $G$ is a subset of G which also forms a group. K is \emph{nontrivial} if it is neither $G$ itself nor $\{e\}$. The \emph{coset} $gK$ is the set $\{gk; k \in K\}$. $g_1K = g_2K$ if $g_2 = g_1k$ for some $k \in K$. If $gK = Kg$ for all $g \in G$ then $K$ is \emph{normal}. For normal $K$ we can form the \emph{quotient group} $G/K$, which is the set of cosets $\{gK; g \in G\}$. Intuitively, we have collapsed all elements of $K$ into the identity. Sometimes we are sloppy and call $G/K$ a coset. If the elements of $G$ are labelled by a discrete set of parameters $G$ is called a \emph{discrete} group. If the elements of $G$ are labelled by a continuous set of parameters $G$ is called a \emph{Lie Group}. A Lie group is \emph{compact} if its parameters span a compact space. For any two groups $G = \{ g_1, g_2,\ldots\}$ and $H = \{h_1,h_2,\ldots\}$, if the $g_i$'s commute with the $h_i$'s we can form the \emph{direct product group} $ G \otimes H = \{(g_i, h_j)\}$ with the multiplication law $(g_k, h_l) (g_m, h_n) = (g_k g_m,h_l h_n)$. If a group does not contain any nontrivial subgroups, i.e. it cannot be written as a direct product of two nontrivial groups, it is called a \emph{simple} group. Note that we often interchange the terms symmetry and group.

An important concept is that of a \emph{representation}. A representation of a group G is defined by the mapping 
\begin{equation*}
R: g \rightarrow D(g)
\end{equation*}
such that the $D(g)$'s satisfy the same multiplication rules as do the g's. We will work exclusively with matrix representations.

\subsection{Lie Groups}
Much of our work is done using Lie Groups. A Lie Group  $G$ is typically defined by a certain specifically selected representation called the \emph{defining} or \emph{fundamental} representation. Working in that fundamental rep, for all the Lie groups we are interested in we can write any element of the group G as 
\begin{equation*}
g(\alpha_1,\alpha_2,\ldots) = e^{-i \alpha_k t_k}
\end{equation*}
where the real $\alpha_k$'s are called the \emph{group parameters}, since they exactly specify an element of the group, and the $t_k$'s are the \emph{generators} of the group G, which are hermitian. The vector space spanned by the generators is called the \emph{Lie Algebra} of the Lie Group G. The number of diagonal generators of a Lie Group is called its \emph{rank}, and a subgroup of $G$ which has the same rank as $G$ is called \emph{maximal}.

In order for the group axioms to be satisfied, the generators of G must satisfy the commutation relations
\begin{equation} \label{eq:commrelns}
[t_n,t_m] = i f_{nmp} t_p
\end{equation}
where the $f_{nmp}$ are the \emph{structure constants} of the Lie Algebra. For the group elements to be hermitian they must be real numbers, and for simple and semi-simple (i.e. no U(1) invariant subgroups) Lie Groups they are completely antisymmetric. We will now look at some important examples of simple Lie Groups.

\subsubsection{SU(N)}
SU(N) is the group of complex unitary $N\times N$ matrices with determinant 1, i.e. 
\begin{equation} \label{eq:SU(N)defn}
SU(N) = \{M;M\in \mathbb{C}^{N\times N}, M M^{\dagger} = 1, \textrm{det}(M) = 1 \}. 
\end{equation}
If we omit the det$(M) = 1$ condition we get U(N), the group of complex unitary matrices with determinant $e^{i \theta}$ for some arbitrary $\theta$. Note that $U(N) = SU(N) \otimes U(1)$. There are $N^2 -1$ generators $\lambda_i$ of SU(N), which for the defining rep are the independent $N \times N$ hermitian matrices. They are usually normalised to satisfy Tr$(\lambda_l \lambda_m) = \sfrac{1}{2} \delta_{lm}$. SU(N) has rank $N-1$.

\subsubsection{SO(N)}
SO(N) is  defined to be the group of real orthogonal $N \times N$ matrices with determinant 1, i.e.
\begin{equation} \label{eq:SO(N)defn}
SO(N) = \{M;M\in \mathbb{R}^{N\times N}, M M^T = 1, \textrm{det}(M) = 1 \}.
\end{equation}
Note that SO(3) is just the group of 3-D spatial rotation matrices. (If we omit the det$(M) = 1$ condition we get O(N), the group of real orthogonal matrices with determinant 1 or -1. SO(N) is the compact subgroup of O(N).)
Any SO(N) matrix can be expressed as $M = e^A$ where A is real and antisymmetric. We can write A as a linear combination of $N(N-1)/2$ antisymmetric matrices denoted by $i J^{ij}$, where the $J^{ij}$ matrices are  completely imaginary and antisymmetric, hence hermitian, and $J^{ij}$ by itself generates a rotation in the $ij$-plane. The $J$'s obey the commutation relation 
\begin{equation} \label{eq:Jcommreln}
[J^{ij},J^{kl}] = i(\delta^{ik}J^{jl} - \delta^{jk}J^{il} + \delta^{jl}J^{ik} - \delta^{il}J^{jk}).
\end{equation}
The $J^{ij}$'s are the generators of $SO(N)$ and we write $M = e^{i \theta^{ij} J^{ij}}$, where the $\theta^{ij}$'s are the group parameters.

\subsection{Representations}
Technically, a group is completely defined by a labelled list of elements and a complete multiplication table. However, in practise this is very unwieldy, and therefore we like to work in \emph{representations} of the group where we can apply some natural operation -- like matrix multiplication -- to the representations of the group elements to get the same behaviour we would were we to apply our -- possibly infinite -- multiplication table. As mentioned before, a representation of a group G is defined by the mapping 
\begin{equation*}
R: g \rightarrow D(g)
\end{equation*}
which preserves multiplicative structure, i.e.
\begin{equation*}
\textrm{if} \qquad g_i g_j = g_k, \qquad \textrm{then} \qquad D(g_i) D(g_j) = D(g_k)
\end{equation*}
The fundamental representations used to define SU(N) and SO(N) are given in Equations \ref{eq:SU(N)defn} and \ref{eq:SO(N)defn} and are very familiar. But since representations are the bread and butter of our work we need to get familiar with larger reps.

\subsubsection{Representations of SU(N)}
An N-element vector $T^i$ transforms under the action of an SU(N) group element $U$ as 
\begin{equation*}
T^i \rightarrow {T^i}' = U^i_a T^a
\end{equation*}
$T^i$ is acted upon by the fundamental rep of SU(N), and we call it $N$, the tensor representation of SU(N) with $N$ independent elements. When we say ``tensor representation", what we mean is that we can derive a representation of SU(N) from the way $N$ transforms. This is trivial here since we used the fundamental rep which defines SU(N) to derive the way $T^i$ transforms, but it will be important later on. The distinction between a tensor and the representation it induces by the way it transforms is too rarely made explicit and a potential source of confusion, since it is the tensor that is called ``the 10" (if it has 10 independent elements, say), but we also refer to it as ``a representation". Having made that point clear, I will now succumb to the nomenclative laziness of my field and call everything under the sun a representation. 

To start building larger reps, we need to define another object $T_i$, which transforms as the complex conjugate of $T^i$ and is labelled $\overline{N}$:
\begin{equation*}
T_i \rightarrow (U^\dagger)_i^a T_a = {U^i_a}^* T_a
\end{equation*}

I define the upper and lower indices of $U$ to label the element's row and column respectively. With this convention, $T^i$ and $T_i$ are column and row vectors respectively which are acted upon by simple matrix multiplication. 

From these two basic objects we can derive many other representations by use of the \emph{tensor product}. If we want to build up a higher-rank tensor, we just say that it \emph{transforms like} a product of $N$'s and $\overline{N}$'s (saying nothing about the actual value of the tensor elements apart from restrictions placed on them by the required transformation properties). For example, we construct a rank (2,1) tensor like this
\begin{equation*}
T^{ij}_k \thicksim T^i T^j T_k
\end{equation*}
(where the $\thicksim$ means ``transforms like") which, under action of SU(N), transforms as
\begin{align*}
T^i T^j T_k &\rightarrow U^i_a U^j_b {U^k_c}^* T^a T^b T_c \\
T^{ij}_k &\rightarrow U^i_a U^j_b {U^k_c}^* T^{ab}_c
\end{align*}
Clearly we can relabel the independent elements $T^{ij}_k$ of that tensor as $V^i$, i.e. simply arranging them in a vector, and rewrite the action of SU(N) as 
\begin{equation*}
V^i \rightarrow M^i_a V^a.
\end{equation*}
Briefly returning to nomenclative pedantry for clarity, M is in actual fact a representation of SU(N) induced by the way the tensor transforms. For each $U \in SU(N)$, we can find a corresponding $M$. These $M$'s clearly follow the same multiplication rules as the $U$'s. 

Tensors of arbitrary rank $(n,m)$ transform as
\begin{equation*}
T^{i_1 i_2 \ldots i_n}_{j_1 j_2 \ldots j_m} \rightarrow U^{i_1}_{a_1} U^{i_2}_{a_2} \ldots U^{i_n}_{a_n} {U^{j_1}_{b_1}}^* {U^{j_2}_{b_2}}^* \ldots {U^{j_m}_{b_m}}^* T^{a_1 a_2 \ldots a_n}_{b_1 b_2 \ldots b_m}
\end{equation*}
For SU(N) one gets the conjugate of a tensor rep by noting that it transforms like the rep's complex conjugate (in index notation). Keeping in line with our established notation, this corresponds to making upper indices lower indices and vice versa. Also, we change the label from $x$ to $\bar{x}$ (or vice versa).

Only irreducible representations are given a label according to their \emph{dimension} (number of independent components). In order to get \emph{irreducible} representations, which are defined as reps which cannot be brought into block-diagonal form by a similarity transformation, we dissect our tensor into components (not talking about the tensor's elements here) with definite symmetry or trace properties which transform into themselves. In order to do that we need to know about those tensors which are \emph{invariant} or \emph{isotropic} under SU(N) transformation. There are only two: the Kroenecker Delta $\delta^i_j$ and the rank-N Levi-Civita Symbols $\epsilon_{i_1 i_2 \ldots i_N}$ and $\epsilon^{i_1 i_2 \ldots i_N}$ (where, just to reiterate, the position of each index indicates whether it transforms with a $U$ or a $U^*$). Technically these two invariant tensors are $1$ representations of SU(N), since they transform like the identity. We can hence, by use of the tensor product, attach them to any tensor we like without introducing ``additional information", which would be akin to a non-trivial tensor product. We also note that tracing over an upper and a lower index is invariant under SU(N) transformation, easily seen from the unitarity of the $U$'s:
\begin{equation*}
R^{\alpha \ldots}_{\ldots} L^{\ldots}_{\alpha \ldots} \rightarrow U^\alpha_a (U^\dagger)^b_\alpha \ldots R^{a \ldots}_{\ldots} L^{\ldots}_{b \ldots} = \delta^a_b R^{a \ldots}_{\ldots} L^{\ldots}_{b \ldots} = R^{a \ldots}_{\ldots} L^{\ldots}_{a \ldots*}
\end{equation*}
Tracing over lower and upper indices of two different tensors is called \emph{contraction}, and we can use the invariant tensors to contract indices of tensors.  Symmetrising and antisymmetrising in SU(N) is only done across two (or more) lower or two (or more) upper indices, not lower and upper, since otherwise the respective symmetry components do not transform into themselves under SU(N) action. If through contraction and (anti)symmetrising we arrive at an object which cannot be contracted or (anti)symmetrised without reducing it to zero, we have ourselves an irreducible representation. An example will serve to illustrate that process.

Say we are working in SU(3) and want to compute the tensor product $3 \otimes 3$. 
\begin{equation*}
3 \otimes 3 = T^{ij} \thicksim T^i T^j
\end{equation*}
Consider the symmetric component
\begin{equation*}
T^{ij}_S = T^{ij} + T^{ji}
\end{equation*}
What happens if we try and contract it with invariant tensors? We know that we can't use the Kroenecker Delta to trace over its two upper indices. How about the Levi-Civita?
\begin{equation*}
A_k = \epsilon_{ijk} T^{ij}_S = 0
\end{equation*}
Therefore the symmetric component is irreducible, and since it has 6 independent components we call it the $6$.\footnote{Note we call $T^S_{ij}$ the $\bar{6}$. In general, if we have tensors with all upper or lower indices we denote the rep with lower indices with the bar. If there are less (more) upper than lower indices we can adopt the practise of (not) having a bar above the labelling number, but that is just convention, and sometimes the bar/no-bar distinction is not enough to distinguish different representations with the same number of independent components. Whenever precise ``indexology" is required I will make the exact form of each rep explicit to avoid confusion.} Now let's play with the asymmetric component
\begin{equation*}
T^{ij}_A = T^{ij} - T^{ji}
\end{equation*}
Once again, we can't mess with the Kroenecker Delta, but we can apply the Levi-Civita without loss of information
\begin{equation*}
B_k = \epsilon_{ijk} T^{ij}_A
\end{equation*}
so for example $B_1 = 2 T^{23}$. We can see that this rep transforms as the $\bar{3}$ of SU(3). Hence $3 \otimes 3 = 6 + \bar{3}$. This example highlights a peculiarity of SU(3): Its $3 \times 3$ antisymmetric matrix rep transforms as a row vector. This is because we can use a rank-$N$ Levi-Civita to contract $(N-1)$ asymmetric indices to one index without loss of information. In general, using the Levi-Civita is only sensible if it allows us to reduce the rank of the tensor - applying a rank $(0,10)$ Levi-Civita to a rank $(3,0)$ tensor would produce a rank $(0,7)$ tensor, with plenty of ``superfluous", meaning trivially non-independent, components -- a step in the wrong direction on the path to irreducibility. 

To display the use of trace here is another SU(3) example. Consider the tensor product $3 \otimes \bar{3}$
\begin{equation*}
3 \otimes \bar{3} = T^i_j \thicksim T^i T_j.
\end{equation*}
We can't do any (anti)symmetrising here, and we can't use Levi-Civita because we don't have enough indices on top or bottom, but we can form the trace $T^\alpha_\alpha$, which is invariant under SU(N) action, i.e. transforms as a singlet $1$. The remaining independent components are
\begin{equation*}
T^{ij} - \sfrac{1}{3} T^\alpha_\alpha
\end{equation*}
obtained by simply removing the trace. This rep is a traceless $3 \times 3$ matrix and has 8 independent components, so we call it the $8$. Hence $3 \otimes \bar{3} = 1 + 8$. (The $8$ also has the special property of being the adjoint rep of SU(3), which is of great importance to particle physics, and we comment on it in Section \ref{sec:sm}.) These two examples are special cases of a useful general result for SU(N):
\begin{align} \label{eq:SU(N)NxN}
N \otimes N &= \sfrac{1}{2}N(N+1)_S + \overline{\sfrac{1}{2}{N(N-1)}}_A \nonumber \\
N \otimes \overline{N} &= 1 + (N^2 -1)
\end{align}

\subsubsection{Representations of SO(N)}
The representations of SO(N) are formed in exactly the same way as those of SU(N), with the following changes:
\begin{itemize}
\item There are no bottom indices since SO(N) matrices are real, and hence all representations are real (with the exception of spinor reps, but never mind that for now, that will be addressed in Section \ref{sec:theory}). 
\item Traces across any two indices are invariant, which means that all higher-rank irreducible representations of SO(N) are traceless
\end{itemize}
We can get a similar result for $N\otimes N$ as for SU(N), taking into account that trace is always invariant:
\begin{equation} \label{eq:SO(N)NxN}
N \otimes N = (\sfrac{1}{2}N(N+1)- 1)_S + 1 + {\sfrac{1}{2}{N(N-1)}}_A
\end{equation}

\subsubsection{The Adjoint Representation}
Let our Lie Group have $d$ generators. By applying the Jacobi Identity to its generators
\begin{equation*}
[t_j,[t_k,t_l]] + [t_l,[t_j,t_k]] + [t_k,[t_l,t_j]] = 0
\end{equation*}
and using Equation \ref{eq:commrelns} we get a relation among the structure constants
\begin{equation*}
f_{mnp}f_{lpq} + f_{lmp}f_{npq} + f_{nlp}f_{mpq} = 0.
\end{equation*}
Hence we can define d  matrices by
\begin{equation*}
(X_l)_{mn} = -i f_{lmn}
\end{equation*}
which satisfy
\begin{equation*}
[X_l,X_m] = i f_{lmp} X_p.
\end{equation*}
These X matrices are a $d \times d$ representation of the Lie Algebra (recall the fundamental representation of the Lie Algebra is $N \times N$) and are closely connected to the irreducible tensor representation $d$ of the Lie Group, called the adjoint representation. To see this let's work through the SU(3) case, which is of special importance for the Standard Model. This procedure easily generalises to SO(N) and SU(N).

SU(3) has $3^2 - 1 = 8$ generators, so its adjoint rep is the $8$, the traceless $T^i_j$, which is also hermitian since it has equal numbers of top and bottom indices. This $3 \times 3$ traceless hermitian matrix can be expressed as a linear combination of SU(3) generators
\begin{equation*}
T^i_j = G_q (\lambda_q)^i_j
\end{equation*}
working in matrix notation to avoid too many indices, note that $T$ transforms as $T \rightarrow U T U^\dagger$. We have $U = e^{i \theta_p \lambda_p}$, and for the moment let's restrict ourselves to an infinitesimal SU(3) transformation $U = 1 + i \theta_p \lambda_p$. Working to first order in the group parameters, write how T transforms under this infinitesimal U and simplify,
\begin{align*}
T &= G_q \lambda_q \nonumber \\
T &\rightarrow U T U^\dagger \nonumber \\
&\phantom{\rightarrow} = (1 + i \theta_p \lambda_p)(G_q \lambda_q)(1 - i \theta_s \lambda_s) \nonumber \\
&\phantom{\rightarrow} = G_q (\lambda_q + i \theta_p [\lambda_p,\lambda_q]) \nonumber \\
&\phantom{\rightarrow} = G_q (\lambda_q + i \theta_p f_{pqr} \lambda_r) \nonumber \\
&\phantom{\rightarrow} = G_r \lambda_r + i \theta_p f_{pqr} G_q \lambda_r \nonumber \\
\nonumber \\
T = G_r \lambda_r &\rightarrow (G_r - i \theta_p f_{prq} G_q) \lambda_r \nonumber \\
&\phantom{\rightarrow} = (G_r + (\theta_p X_p)_{rq}  G_q) \lambda_r \nonumber \\
\end{align*}
where I renamed some dummy indices. We can write this as
\begin{equation*}
V = (1 +  \theta_p X_p ) V
\end{equation*}
where 
\begin{equation*}
V = \left( \begin{array}{c} G_1 \\ G_2 \\ \vdots \\ G_8 \end{array} \right).
\end{equation*}
is the vector in $\lambda$-basis representing $T^i_j$. For finite SU(3) transformations, this turns into 
\begin{equation*}
V = e^{\theta_p X_p} V.
\end{equation*}
This process is trivially generalisable to SU(N) and also works for SO(N), where the adjoint rep is the $\sfrac{1}{2}N(N-1)$, the rank-2 antisymmetric tensor.

The connection is now clear: the matrix representation determined by the way the adjoint tensor representation transforms is generated by the X matrices, which are a $d \times d$ representation of the Lie Algebra.

\section{Useful subgroups of SO(10)}
\label{ap:embeddings}
Here we will discuss in detail how the relevant subgroups of SO(10) are embedded for the asymmetric Higgs kink scenario discussed in \ref{subsubsec:edwards}.

\subsection{The Embedding of $U(5)$ in $SO(10)$}

There is a simple mapping, call it $h$, which takes any complex number to a real $2 \times 2$ matrix.
\begin{displaymath}
r e^{i\theta} \stackrel{h}{\longrightarrow}
r \left(\begin{array}{cc}
\cos{\theta} & -\sin{\theta}\\
\sin{\theta} & \cos{\theta} \end{array} \right) \qquad r \geq 0
\end{displaymath}
$h$ preserves both additive and multiplicative structure. This function can be generalised to map any $n \times n$ complex matrix to a $2n \times 2n$ real matrix, where each complex entry is mapped to the corresponding $2 \times 2$ sub-block of the real matrix.

We shall now use our generalised $h$ to show how $U(5)$ can be embedded in $SO(10)$. For any complex matrix $M$ first note that
\begin{equation*}
{h}(M^\dagger) = {h}(M)^{T} \nonumber
\end{equation*}
Let $U \in U(5)$. Since $h$ preserves both additive and multiplicative structure it is easy to see that
\begin{align*}
{h}(U^\dagger U) &= {h}(\boldsymbol{1}_5) = \boldsymbol{1}_{10} \nonumber \\
&= {h}(U)^{T} {h}(U) \\
\\
{h}(U_{1} U_{2}) &= {h}(U_{1}){h}(U_{2})\textrm{.}
\end{align*}
Hence $h$ maps $U(5)$ into $O(10)$. But $U(5)$ is a connected manifold, and since the antispecial $O(10)$ matrices do not form a subgroup, $h$ maps $U(5)$ into $SO(10)$.

The restriction of $SO(10)$ to $U(5)$ is therefore realised by restricting the choice of possible $SO(10)$ transformation matrices to those which are also a member of the image of $h$. Consider an arbitrary $U(5)$ matrix $U$ with entries $U_l^k = r_l^k e^{i \theta_l^k}$, i.e. 
\begin{displaymath}
U = \left( \begin{array}{ccc}
r_1^1 e^{i \theta_1^1} & r_2^1 e^{i \theta_2^1} & \ldots \\
r_1^2 e^{i \theta_1^2} & r_2^2 e^{i \theta_2^2} & \ldots \\
\vdots & \vdots & \ddots
\end{array} \right)
\end{displaymath}
The corresponding SO(10) matrix is
\begin{displaymath}
{h}(U) = \left( \begin{array}{ccc}
r_1^1\left(\begin{array}{cc} 
\cos{\theta_1^1} & -\sin{\theta_1^1}\\ 
\sin{\theta_1^1} & \cos{\theta_1^1} \end{array} \right) & 
r_2^1\left(\begin{array}{cc} 
\cos{\theta_2^1} & -\sin{\theta_2^1}\\ 
\sin{\theta_2^1} & \cos{\theta_2^1} \end{array} \right) & 
\ldots \\
 & & \\
r_2^1\left(\begin{array}{cc} 
\cos{\theta_2^1} & -\sin{\theta_2^1}\\ 
\sin{\theta_2^1} & \cos{\theta_2^1} \end{array} \right) & 
r_2^2\left(\begin{array}{cc} 
\cos{\theta_2^2} & -\sin{\theta_2^2}\\ 
\sin{\theta_2^2} & \cos{\theta_2^2} \end{array} \right) & 
\ldots \\
 & & \\
\vdots & \vdots & \ddots\\
\end{array} \right)
\end{displaymath}
Upon careful inspection one realises that this restricted $SO(10)$ matrix can be expressed in index notation as
\begin{equation} \label{eq:restrictedO}
O^i_j = r^k_l \Big[ \delta^i_{2k-1}(\delta^{2l-1}_j\cos{\theta^k_l} - \delta^{2l}_j \sin{\theta^k_l}) +  \delta^i_{2k}(\delta^{2l}_j\cos{\theta^k_l} + \delta^{2l-1}_j \sin{\theta^k_l}) \Big]  
\end{equation}

\subsection{The Embedding of $U(3) \otimes U(2)$ in $U(5)$}
$U(5)$ is equivalent to $SU(5) \otimes U(1)'$. We can further break down the $SU(5)$ into $SU(3) \otimes SU(2) \otimes U(1)''$ by trivially embedding the respective $3 \times 3$ and $2 \times 2$ transformation matrices in the $5 \times 5$ $U(5)$ matrix in a variety of ways. In order to be consistent with the specific $U(3) \otimes U(2)$ symmetry obeyed by the asymmetric kink in the bulk (see \cite{edwards} and \ref{subsubsec:edwards}), we construct the embedding as follows
\begin{equation} \label{eq:restrictedU}
U = \left(\begin{array}{cc}
U_3 & 0\\
0 & U_2 \end{array} \right)
\end{equation}
for $U \in SU(5)$. The $U_3$ and $U_2$ sub-matrices are $SU(3)$ and $SU(2)$ transformation matrices which carry a $U(1)''$ charge of -2 and 3 respectively.

\section{Aside: Direct Products of Spinor Representations}
\label{ap:spinordirectproduct}
We had to teach ourselves how to perform direct products with Spinors. This Appendix does not represent a rigorous analysis, but rather a very informal documentation of our learning process on how to do direct products with spinor representations.

We work in SO(10). From our analysis in \ref{subsec:16}, we know that 
\begin{equation*}
\Psi_A, \overline{\Psi}_B \thicksim \overline{16}, \qquad \qquad \overline{\Psi}_A, \Psi_B \thicksim 16,
\end{equation*}
where
\begin{equation*}
\Psi_A = \sfrac{1}{2}(1-\gamma^\textrm{FIVE})\Psi, \qquad \Psi_B = \sfrac{1}{2}(1+\gamma^\textrm{FIVE})\Psi, \qquad \overline\Psi_A = \sfrac{1}{2} \overline\Psi (1-\gamma^\textrm{FIVE}), \qquad \overline{\Psi}_B = \sfrac{1}{2} \overline\Psi(1+\gamma^\textrm{FIVE}).
\end{equation*}
The key point in performing spinor direct products is that the \emph{matrix indices} of the $32 \times 32$ Clifford Algebra matrices run from 1 to 10 and hence bridge the gap between the SO(10) tensor indices, which run from 1 to 10, and the SO(10) spinor indices, which run from 1 to 32.

We perform the $16 \otimes 16$ product as follows: Start by simply multiplying together two $16$'s in appropriate row/column vector form and inserting the maximum number of $\gamma$'s between them:
\begin{equation*}
\overline{\Psi}_A\gamma \gamma \gamma \gamma \gamma \Psi_B.
\end{equation*}
We have to insert \emph{odd} numbers of $\gamma$'s since if we have an even number of $\gamma$'s between the two Spinors, we would get a factor 
\begin{equation*}
(1 + \gamma^\textrm{FIVE})(1 - \gamma^\textrm{FIVE}) = 0
\end{equation*}
The maximum number of $\gamma$'s we can put in is 5. If we had, say, 7 $\gamma$'s, we could insert a $\gamma^\textrm{FIVE}$ next to a spinor, which is equivalent to a trivial overall change in sign since the Spinors are $\gamma^\textrm{FIVE}$ eigenstates, turning the 7 $\gamma$'s into 3 $\gamma$'s. In a sense, the $\gamma^\textrm{FIVE}$ is a spinor direct product version of the Levi-Civita, in terms of how it is used to obtain irreducible representations. 
	
Now to actually get the product representations, we simply trace over different numbers of $\gamma$'s and remove the remaining traces:
\begin{itemize}
\item $\overline{\Psi}_A \gamma^i \Psi_B$  is a 10 and obtained by tracing over 4 $\gamma$'s. 
\item $\overline{\Psi}_A\gamma^i \gamma^j \gamma^k\Psi_B - \{\textrm{traces}\}$ is a 120, the rank-3 antisymmetric tensor, and is obtained by tracing over 2 $\gamma$'s
\item $\overline{\Psi}_A\gamma^i \gamma^j \gamma^k \gamma^l \gamma^m \Psi_B - \{\textrm{traces}\}$ is a rank-5 antisymmetric tensor, which normally has 252 components. However, since (for example) $\gamma^1 \gamma^2 \gamma^3 \gamma^4 \gamma^5$ can be obtained from $\gamma^6 \gamma^7 \gamma^8 \gamma^9 \gamma^{10}$ by multiplying it by a $\gamma^\textrm{FIVE}$, half of the entries are not independent: the representation is the 126.
\end{itemize}
Hence $16 \otimes 16 = 10 \oplus 120 \oplus 126$. This gives us the form of the $126$ in terms of the basic spinor reps it is derived from. Note that the $\overline{126}$ is obtained by switching $A$ and $B$ in the above computation. 

For completeness we also outline computation $\overline{16} \otimes 16$ direct product, which is performed using exactly the same principles. Multiply together a $16$ and a $\overline{16}$ in appropriate row/column vector form and insert the maximum number of $\gamma$'s between them:
\begin{equation*}
\overline{\Psi}_B\gamma \gamma \gamma \gamma \Psi_B.
\end{equation*}
We have to insert \emph{even} numbers of $\gamma$'s and the maximum number if 4, for the same reasons as for the $16 \otimes 16$. Upon tracing over different numbers of $\gamma$'s and removing remaining traces we obtain the product representations:
\begin{itemize}
\item $\overline{\Psi}_B \Psi_B$ is a singlet and obtained by tracing over 4 $\gamma$'s. 
\item $\overline{\Psi}_B\gamma^i \gamma^j \Psi_B - \{\textrm{traces}\}$ is a 45, the rank-2 antisymmetric tensor, and is obtained by tracing over 2 $\gamma$'s
\item $\overline{\Psi}_A\gamma^i \gamma^j \gamma^k \gamma^l \Psi_B - \{\textrm{traces}\}$ is the 210, the rank-4 antisymmetric tensor, and is obtained by tracing over 4 $\gamma$'s
\end{itemize}
Hence $\overline{16} \otimes 16 = 1 \oplus 45 \oplus 210$.

\section{Calculations for the 120}
\label{ap:calc120}
Here we present the details of calculating the breakdown of the $120$ of SO(10) under restriction to $G_{SM} \otimes U(10$, embedded to conform with the left-over symmetry of the asymmetric kink in the 45 representation. We have to compute the breakdowns of the 10 and 45 first, since the 120 can be obtained from the $10 \otimes 45$ direct product, and the 45 can be obtained from the $10 \otimes 10$ direct product.

\subsection{The 10 Breakdown}
The 10 of $SO(10)$ is a column-vector with 10 real components $T^i = (T^1, T^2, \ldots, T^{10})$, which transforms according to 
\begin{displaymath}
T^i \longrightarrow O^i_a T^a
\end{displaymath}
Under the restricted O given by Equation \ref{eq:restrictedO}, the components can be grouped into two classes based on the way they transform under the restriction of $SO(10)$ to $U(5)$:
\begin{align}
T^{2k-1} &\longrightarrow r^k_l(\cos{\theta^k_l} T^{2l-1} - \sin{\theta^k_l}T^{2l}) \nonumber \\
T^{2k\phantom{-1}} &\longrightarrow r^k_l (\sin{\theta^k_l}T^{2l-1} + \cos{\theta^k_l} T^{2l}) \label{10transform}
\end{align}
where k runs from 1 to 5. We now search for embedded $U(5)$ representations. 

One would expect the 10 of $SO(10)$ to break down into a $5$ and a $\bar{5}$ (since the 10 is real), which turns out to be the case. The 5 of $SU(5)$ (we will deal with the $U(1)'$ charge later) is a column-vector with 5 complex components $\tilde{T}^i = (\tilde{T}^1, \tilde{T}^2, \ldots ,\tilde{T}^5)^T$. Similarly the $\bar{5}$ is a row-vector with 5 complex components $\tilde{T}_i = (\tilde{T}_1, \tilde{T}_2, \ldots ,\tilde{T}_5)$. Under an $SU(5)$ transformation corresponding to the restricted $SO(10)$ transformation $O$, the 5 transforms as 
\begin{align*}
\tilde{T}^k \longrightarrow & \, U^k_l \tilde{T}^l = r^k_l e^{i\theta^k_l} \tilde{T}^l  \\
{Re}\tilde{T}^k\longrightarrow  & \, r^k_l(\cos{\theta^k_l}{Re}\tilde{T}^l - sin{\theta^k_l}{Im}\tilde{T}^l)  \\
{Im}\tilde{T}^k \longrightarrow & \, r^k_l(\sin{\theta^k_l}{Re}\tilde{T}^l + cos{\theta^k_l}{Im}\tilde{T}^l) 
\end{align*}
The embedding of the $SU(5)$ 5 in the $SO(10)$ 10 is now obvious by direct comparison with Equation \ref{10transform}:
\begin{displaymath}
\tilde{T}^k = T^{2k-1} + i T^{2k}
\end{displaymath}
Similarly for the $\bar{5}$. It transforms as
\begin{align*}
\tilde{T}_k \longrightarrow & \, (U^{\dagger})^l_k \tilde{T}_l = r^k_l e^{-i\theta^k_l} \tilde{T}_l  \\
{Re}\tilde{T}_k\longrightarrow  & \, r^k_l(\cos{\theta^k_l}{Re}\tilde{T}_l - sin{\theta^k_l}{Im}\tilde{T}_l)  \\
{Im}\tilde{T}_k \longrightarrow & \, -r^k_l(\sin{\theta^k_l}{Re}\tilde{T}_l + cos{\theta^k_l}{Im}\tilde{T}_l) 
\end{align*} 
and is embedded via
\begin{displaymath}
\tilde{T}_k = T^{2k-1} - i T^{2k}
\end{displaymath}

As for the $U(1)'$ quantum numbers, they must be equal and opposite since the 10 is real. We can normalise them any way we like and choose the value 2. 

I will now introduce a notation which is useful for working with large breakdowns. Under each representation I write how it is embedded in the mother-representation. Hence I write the branching rule for the 10 in \\ $SO(10) \supset SU(5) \otimes U(1)'$ as
\begin{equation}
\begin{array}{ccccc}\label{eq:prelim10breakdown}
10 & \longrightarrow & 5(2) & \oplus & \bar{5}(-2) \\
\scriptstyle T^i & &\scriptstyle \tilde{T}^k = T^{2k-1} + i T^{2k} & &\scriptstyle  \tilde{T}_k = T^{2k-1} - i T^{2k}
\end{array}
\end{equation}

But we can't stop there. Restricting ourselves to the embedded $SU(3) \otimes SU(2) \otimes U(1)''$ subgroup of the $SU(5)$, we see that a 5 of $SU(5)$ breaks down trivially as 
\begin{equation}
\begin{array}{ccccc}\label{eq:5breakdown}
5 & \longrightarrow & (3,1)(-2) & \oplus & (1,2)(3) \\
\scriptstyle \tilde{T}^i & & \scriptstyle A^{\rho} = \tilde{T}^{\rho} & &\scriptstyle  A^{\nu} = \tilde{T}^{\nu + 3}
\end{array}
\end{equation}
i.e. the first 3 entries transform as a 3 of $SU(3)$ and the last 2 entries transform as a 2 of $SU(2)$.\footnote{The 2 and the $\bar{2}$ of $SU(2)$ are not independent of each other, they are related via $B^{\nu} = \epsilon^{\nu a} B_a$. Whenever the conjugate representation of a 2 turns up I will leave it as a $\bar{2}$ to simplify the calculations, and only for the final important results will I rewrite it to a 2. Whenever there is a rank-2 Levi-Cevita it is usually a smoking gun for writing a $\bar{2}$ as a $2$.}

Combining Equations \ref{eq:prelim10breakdown} and \ref{eq:5breakdown} we see that the branching rule for \\ $SO(10) \supset SU(3) \otimes SU(2) \otimes U(1)'' \otimes U(1)'$ is
\begin{equation}
\begin{array}{cccccc}\label{eq:10breakdown}
10 & \longrightarrow & \stackrel{I}{(1,2)(3)(2)} & \oplus & \stackrel{II}{(3,1)(-2)(2)} & \oplus  \\
\scriptstyle T^i & &\scriptstyle A^{\nu} = T^{2\nu+5} + i T^{2\nu + 6} & & \scriptstyle A^{\rho} = T^{2\rho-1} + i T^{2\rho} & \\
\\
 & & \stackrel{III}{(1,\bar{2})(-3)(-2)} & \oplus & \stackrel{IV}{(\bar{3},1)(2)(-2)} & \\
 & & \scriptstyle A_{\nu} = T^{2\nu+5} - i T^{2\nu + 6} & & \scriptstyle A_{\rho} = T^{2\rho-1} - i T^{2\rho} &
\end{array}
\end{equation}
The roman numerals above the representations are labels which we will need later when evaluating the 120 breakdown.

\subsection{The 45 Breakdown}
The 45 of $SO(10)$ is $T^{ij}_A$, a $10 \times 10$ antisymmetric matrix (omit the A for the rest of this Section to avoid index poisoning, as far as that is possible \ldots). It is constructed from the 10 via $10 \otimes_A 10$. For the purpose of deriving transformation properties \emph{only}, write $T^{ij} =  T^i T^j$ (where antisymmetry between i and j indices is implied), and similarly with the breakdown representations of the 45 as they are derived from the breakdown representations of the 10. 

This is the tensor product of the breakdown representations we have to compute:
\begin{equation}\label{eq:45product}
\begin{array}{ccccccccl}
45 & = & 10 \otimes_A 10 & \rightarrow & \Big[ & (3,1)(-2)(2) & \oplus & (1,2)(3)(2) & \oplus \\
\scriptstyle T^{ij} & & & & & \scriptstyle A^{\rho} = T^{2\rho-1} + i T^{2\rho} & & \scriptstyle A^{\nu} = T^{2\nu+5} + i T^{2\nu + 6} & \\
 & & & & & (\bar{3},1)(2)(-2) & \oplus & (1,\bar{2})(-3)(-2)& \Big] \otimes_A \Big[ \ldots \Big]   \\
 & & & & & \scriptstyle A_{\rho} = T^{2\rho-1} - i T^{2\rho} & & \scriptstyle A_{\nu} = T^{2\nu+5} - i T^{2\nu + 6} & \\
\end{array}
\end{equation}

We then work out each of the terms in the above tensor product. Because of the antisymmetry of the 45, terms like $(1,2)\otimes_A(3,1)$ and $(3,1)\otimes_A(1,2)$ are not independent, and purely symmetric products are zero. A term-by-term breakdown of the left-hand-side of Equation \ref{eq:45product} is as follows (irreducible terms are in bold):

\begin{itemize}

\item $(1,2\otimes 2)(6)(4)_A$:
\begin{align*}
B^{\nu \mu}_A &= \big[ (T^{2\nu + 5}T^{2\mu + 5} - T^{2\nu + 6}T^{2\mu + 6}) + i(T^{2\nu+5}T^{2\mu+6}+T^{2\nu+6}T^{2\mu + 5}) \big]_A  \\
 &= \big[ (T^{2\nu + 5,2\mu + 5} - T^{2\nu + 6,2\mu + 6}) + i(T^{2\nu+5,2\mu+6}+T^{2\nu+6,2\mu + 5}) \big]_A 
\end{align*}
The antisymmetric component is the singlet $B^{1,2}$.
\begin{displaymath}
\boldsymbol{(1,1)(6)(4)}: B = (T^{7,9} - T^{8,10}) + i(T^{7,10} + T^{8,9})
\end{displaymath}
(From now on I will skip the step of writing the two separate $T^i$s)

\item $\boldsymbol{(3,2)(1)(4)}$:
\begin{displaymath}
B^{\rho \nu} = (T^{2\rho-1,2\nu+5} - T^{2\rho, 2\nu+6}) + i(T^{2\rho,2\nu+5} + T^{2\rho-1, 2\nu+6})
\end{displaymath}

\item $(1,\bar{2} \otimes 2)(0)(0)$:
\begin{displaymath}
B^{\nu}_{\mu} = (T^{2\nu+5,2\mu+5} + T^{2\nu + 6, 2\mu+6}) + i(T^{2\nu + 5,2\mu+6} + T^{2\nu + 6, 2\mu+5})
\end{displaymath}
gives
\begin{align*}
\boldsymbol{(1,1)(0)(0)} \textrm{: } B\phantom{^{\nu}_{\mu}} = \; &T^{2\alpha + 5, 2\alpha + 6} \\
\boldsymbol{(1,3)(0)(0)} \textrm{: } B^{\nu}_{\mu} = \; &(T^{2\nu+5,2\mu+5} + T^{2\nu + 6, 2\mu+6}) + \\
& i(T^{2\nu + 5,2\mu+6} + T^{2\nu + 6, 2\mu+5} - \delta^{\nu}_{\mu} T^{2\alpha + 5, 2\alpha + 6}) 
\end{align*}

\item $\boldsymbol{(\bar{3},2)(5)(0)}$: 
\begin{displaymath}
B^{\nu}_{\rho} = (T^{2\rho-1,2\nu+5} + T^{2\rho, 2\nu+6}) - i(T^{2\rho,2\nu+5} - T^{2\rho-1, 2\nu+6})
\end{displaymath}

\item $(3\otimes 3,1)(-4)(4)$:
\begin{displaymath}
B^{\rho \sigma} = (T^{2\rho-1,2\sigma - 1} -T^{2\rho, 2\sigma}) + i(T^{2\rho-1,2\sigma} + T^{2\rho, 2\sigma -1})
\end{displaymath}
gives 
\begin{displaymath}
\boldsymbol{(\bar{3},1)(-4)(4)}: B_{\rho} = \epsilon_{\rho a b} \big[ (T^{2a-1,2b - 1} -T^{2a, 2b}) + i(T^{2a-1,2b} + T^{2a, 2b -1}) \big]
\end{displaymath}

\item $\boldsymbol{(3,\bar{2})(-5)(0)}$:
\begin{equation*}
B_{\nu}^{\rho} = (T^{2\rho-1,2\nu+5} + T^{2\rho, 2\nu+6}) + i(T^{2\rho,2\nu+5} - T^{2\rho-1, 2\nu+6})
\end{equation*}

\item $(3\otimes \bar{3},1)(0)(0)$:
\begin{equation*}
B^{\rho}_{\sigma} = (T^{2\rho-1,2\sigma - 1} + T^{2\rho, 2\sigma}) + i(T^{2\rho,2\sigma-1} - T^{2\rho-1, 2\sigma})
\end{equation*}
gives
\begin{align*}
\boldsymbol{(1,1)(0)(0)}\textrm{: } B\phantom{^{\rho}_{\sigma}} = \; &T^{2\beta, 2\beta -1} \\
\boldsymbol{(8,1)(0)(0)}\textrm{: } B^{\rho}_{\sigma} = \; &(T^{2\rho-1,2\sigma - 1} + T^{2\rho, 2\sigma}) + \\
& i(T^{2\rho,2\sigma-1} - T^{2\rho-1, 2\sigma} - \sfrac{2}{3}\delta^{\rho}_{\sigma} T^{2\beta, 2\beta -1})
\end{align*}

\item $(1,\bar{2}\otimes \bar{2})(-6)(-4)$:
\begin{equation*}
{B_{\nu \mu}}_A = \big[ (T^{2\nu + 5,2\mu + 5} - T^{2\nu + 6,2\mu + 6}) + i(T^{2\nu+5,2\mu+6}+T^{2\nu+6,2\mu + 5}) \big]_A 
\end{equation*}
gives
\begin{equation*}
\boldsymbol{(1,1)(-6)(-4)}: B = (T^{7,9} - T^{8,10}) - i(T^{7,10} + T^{8,9})
\end{equation*}

\item $\boldsymbol{(\bar{3},\bar{2})(-1)(-4)}$:
\begin{equation*}
B_{\rho \nu} = (T^{2\rho-1,2\nu+5} - T^{2\rho, 2\nu+6}) - i(T^{2\rho,2\nu+5} + T^{2\rho-1, 2\nu+6})
\end{equation*}

\item $(\bar{3}\otimes\bar{3},1)(4)(-4)_A$:
\begin{displaymath}
B_{\rho \sigma} = (T^{2\rho-1,2\sigma - 1} -T^{2\rho, 2\sigma}) - i(T^{2\rho-1,2\sigma} + T^{2\rho, 2\sigma -1})
\end{displaymath}
gives 
\begin{displaymath}
\boldsymbol{(3,1)(4)(-4)}: B^{\rho} = \epsilon^{\rho a b} \big[ (T^{2a-1,2b - 1} -T^{2a, 2b}) - i(T^{2a-1,2b} + T^{2a, 2b -1}) \big]
\end{displaymath}
\end{itemize}

We have just derived the complete breakdown of the 45. Let us write it out once in all its glory, in a vertical form of our usual notation because of the sheer size of the thing. The numbers in curly brackets are labels for each of the breakdown representations which we will need when evaluating the 120 breakdown.

\begin{equation}\label{eq:45breakdown}
\begin{array}{c l|l}
\qquad \! \! 45 & & T^{ij}\\
& \\
\qquad \downarrow & \\
& \\
\{1\} & (1,1)(0)(0) & B = T^{2\alpha + 5, 2\alpha + 6} \\
&\qquad \oplus\\
\{2\} & (1,1)(0)(0) & B = T^{2\beta, 2\beta -1}\\
&\qquad \oplus\\
\{3\} & (1,1)(6)(4) & B = (T^{7,9} - T^{8,10}) + i(T^{7,10} + T^{8,9})\\
&\qquad \oplus\\
\{4\} & (1,1)(-6)(-4) & B = (T^{7,9} - T^{8,10}) - i(T^{7,10} + T^{8,9})\\
&\qquad \oplus\\
\{5\} & (\bar{3},1)(-4)(4) & B_{\rho} = \epsilon_{\rho a b} \big[ (T^{2a-1,2b - 1} -T^{2a, 2b}) + i(T^{2a-1,2b} + T^{2a, 2b -1}) \big]\\
&\qquad \oplus\\
\{6\} &(3,1)(4)(-4) & B^{\rho} = \epsilon^{\rho a b} \big[ (T^{2a-1,2b - 1} -T^{2a, 2b}) - i(T^{2a-1,2b} + T^{2a, 2b -1}) \big]\\
&\qquad \oplus\\
\{7\} &(3,2)(1)(4) & B^{\rho \nu} = (T^{2\rho-1,2\nu+5} - T^{2\rho, 2\nu+6}) + i(T^{2\rho,2\nu+5} + T^{2\rho-1, 2\nu+6})\\
&\qquad \oplus\\
\{8\} &(\bar{3},\bar{2})(-1)(-4) & B_{\rho \nu} = (T^{2\rho-1,2\nu+5} - T^{2\rho, 2\nu+6}) - i(T^{2\rho,2\nu+5} + T^{2\rho-1, 2\nu+6}) \\
&\qquad \oplus\\
\{9\} &(3,\bar{2})(-5)(0) & B_{\nu}^{\rho} = (T^{2\rho-1,2\nu+5} + T^{2\rho, 2\nu+6}) + i(T^{2\rho,2\nu+5} - T^{2\rho-1, 2\nu+6}) \\
&\qquad \oplus\\
\{10\} &(\bar{3},2)(5)(0) & B^{\nu}_{\rho} = (T^{2\rho-1,2\nu+5} + T^{2\rho, 2\nu+6}) - i(T^{2\rho,2\nu+5} - T^{2\rho-1, 2\nu+6}) \\
&\qquad \oplus\\
\{11\} &(1,3)(0)(0) &  B^{\nu}_{\mu} = (T^{2\nu+5,2\mu+5} + T^{2\nu + 6, 2\mu+6}) +\\
& & \phantom{B^{\nu}_{\mu} = (} i(T^{2\nu + 5,2\mu+6} +T^{2\nu + 6, 2\mu+5} - \delta^{\nu}_{\mu} T^{2\alpha + 5, 2\alpha + 6}) \\
&\qquad \oplus\\
\{12\} &(8,1)(0)(0) &  B^{\rho}_{\sigma} = (T^{2\rho-1,2\sigma - 1} + T^{2\rho, 2\sigma}) + \\
& & \phantom{B^{\rho}_{\sigma} = }+ i(T^{2\rho,2\sigma-1} - T^{2\rho-1, 2\sigma} - \sfrac{2}{3}\delta^{\rho}_{\sigma} T^{2\beta, 2\beta -1})\\

\end{array}
\end{equation}

\subsection{The 120 Breakdown}
\label{ap:120breakdown}
We will follow exactly the same procedure as we did for the 45. The 120 of $SO(10)$ is $T^{ijk}_A$, a completely antisymmetric rank-3 tensor (again omit the A from now on) and is obtained from $10 \otimes_A 10 \otimes_A 10$. Its breakdown can be obtained from the breakdown of $45 \otimes 10$ by imposing the restriction that the three indices of the mother-representation are completely antisymmetric. As we will see, this restriction reduces the number of resulting independent non-zero representations from 48 to 24. For the purpose of transformation properties \emph{only}, write $T^{ijk} = T^{ij}T^k$, where antisymmetry between all indices is implied. 

Now we get to use the labels we gave to the 10 and 45 breakdown representations. The tensor product we have to evaluate is:
\begin{equation} \label{eq:120product}
\begin{array}{l}
120 = (45 \otimes 10)_A \longrightarrow \\ \\
\begin{array}{rccccccccc}
\Big[ & (1,1)(0)(0) & \oplus & (1,1)(0)(0) & \oplus & (1,1)(6)(4) & \oplus & (1,1)(-6)(-4) & \oplus\\
 & \scriptstyle{\{1\}} & & \scriptstyle{ \{2\}} & &\scriptstyle{\{3\}} & & \scriptstyle{\{4\}} \\ \\
 & (\bar{3},1)(4)(-4) & \oplus & (3,1)(4)(-4) & \oplus & (3,2)(1)(4) & \oplus & (\bar{3},\bar{2})(-1)(-4) & \oplus\\
 & \scriptstyle{\{5\}} & & \scriptstyle{\{6\}} & &\scriptstyle{\{7\}} & & \scriptstyle{\{8\}} \\ \\
 & (3,\bar{2})(-5)(0) & \oplus & (\bar{3},2)(5)(0) & \oplus & (1,3)(0)(0) & \oplus & (8,1)(0)(0) & \Big] & \bigotimes_A \\
 & \scriptstyle{\{9\}} & & \scriptstyle{\{10\}} & &\scriptstyle{\{11\}} & & \scriptstyle{\{12\}} \\ \\
\Big[ &  (1,2)(3)(2) & \oplus & (3,1)(-2)(2) & \oplus & (1,\bar{2})(-3)(-2) & \oplus & (\bar{3},1)(2)(-2) & \Big] \\
 & \scriptstyle{I} & & \scriptstyle{II} & &\scriptstyle{III} & & \scriptstyle{IV}
\end{array}
\end{array}
\end{equation}
The independent non-zero terms on the LHS are (again irreducible representations in bold):

\begin{itemize}

\item \cross{1}{I} gives
\begin{align*}
(1,2)(3)(2): C^\nu & = T^{2\alpha + 5, 2\alpha + 6}T^{2\nu + 5} + i T^{2\alpha + 5, 2\alpha + 6}T^{2\nu + 6}\\
\textrm{i.e.} \qquad \qquad \qquad \qquad & \\
\boldsymbol{(1,2)(3)(2)}: C^\nu &= T^{2\nu + 5, 2\alpha + 5, 2\alpha + 6} + i T^{2\nu + 6, 2\alpha + 5, 2\alpha + 6}
\end{align*}
From now on I will skip the intermediate step of writing the two separate T's and instead write everything in terms of the $T^{ijk}$ right away.

\item \cross{1}{II} gives
\begin{equation*}
\boldsymbol{(3,1)(-2)(2)}: C^\rho = T^{2\rho - 1, 2\alpha + 5, 2\alpha + 6} + i T^{2\rho, 2\alpha + 5, 2\alpha + 6}
\end{equation*}

\item \cross{1}{III} gives
\begin{align*}
(1,\bar{2})(-3)(-2): C_\nu & = T^{2\nu + 5, 2\alpha + 5, 2\alpha + 6} - i T^{2\nu + 6, 2\alpha + 5, 2\alpha + 6}\\
\textrm{or} \qquad \qquad \qquad \qquad & \\
\boldsymbol{(1,2)(-3)(-2)}: C^\nu &= \epsilon^{\nu a} (T^{2a + 5, 2\alpha + 5, 2\alpha + 6} - i T^{2a + 6, 2\alpha + 5, 2\alpha + 6})
\end{align*}
From now on I will skip the step of first writing down the $\bar{2}$ and write down the $2$ right away.

\item \cross{1}{IV} gives
\begin{equation*}
\boldsymbol{(\bar{3},1)(2)(-2)}: C_\rho = T^{2\rho - 1, 2\alpha +5, 2\alpha+6} - i T^{2\rho, 2\alpha +5, 2\alpha+6}
\end{equation*}

\item \cross{2}{I} gives
\begin{equation*}
\boldsymbol{(1,2)(3)(2)}: C^\nu = T^{2\nu + 5, 2\beta-1, 2\beta} + i T^{2\nu + 6, 2\beta-1, 2\beta}
\end{equation*}

\item \cross{2}{II} gives
\begin{equation*}
\boldsymbol{(3,1)(-2)(2)}: C^\rho = T^{2\rho - 1, 2\beta-1, 2\beta} + i T^{2\rho, 2\beta-1, 2\beta}
\end{equation*}

\item \cross{2}{III} gives
\begin{equation*}
\boldsymbol{(1,2)(-3)(-2)}: C^\nu = \epsilon^{\nu a} (T^{2a + 5, 2\beta-1, 2\beta} - i T^{2a + 6, 2\beta-1, 2\beta})
\end{equation*}

\item \cross{2}{IV} gives
\begin{equation*}
\boldsymbol{(\bar{3},1)(2)(-2)}: C_\rho = T^{2\rho - 1, 2\beta-1, 2\beta} - i T^{2\rho, 2\beta-1, 2\beta}
\end{equation*}

\item \cross{3}{II} gives
\begin{align*}
\boldsymbol{(3,1)(4)(6)}: C^{\rho} = &(T^{2\rho - 1,7,9} - T^{2\rho - 1,8,10} - T^{2\rho,7,10} - T^{2\rho,8,9}) + \\
&i(T^{2\rho,7,9} - T^{2\rho,8,10} + T^{2\rho-1,7,10} + T^{2\rho-1,8,9})
\end{align*}

\item \cross{3}{IV} gives
\begin{align*}
\boldsymbol{(\bar{3},1)(8)(2)}: C_{\rho} = &(T^{2\rho - 1,7,9} - T^{2\rho - 1,8,10} + T^{2\rho,7,10} + T^{2\rho,8,9}) + \\
&i(-T^{2\rho,7,9} + T^{2\rho,8,10} + T^{2\rho-1,7,10} + T^{2\rho-1,8,9})
\end{align*}

\item \cross{4}{II} gives
\begin{align*}
\boldsymbol{(3,1)(-8)(-2)}: C^{\rho} = &(T^{2\rho - 1,7,9} - T^{2\rho - 1,8,10} + T^{2\rho,7,10} + T^{2\rho,8,9}) - \\
&i(-T^{2\rho,7,9} + T^{2\rho,8,10} + T^{2\rho-1,7,10} + T^{2\rho-1,8,9})
\end{align*}

\item \cross{4}{IV} gives
\begin{align*}
\boldsymbol{(\bar{3},1)(-4)(-6)}: C_{\rho} = &(T^{2\rho - 1,7,9} - T^{2\rho - 1,8,10} - T^{2\rho,7,10} - T^{2\rho,8,9}) - \\
&i(T^{2\rho,7,9} - T^{2\rho,8,10} + T^{2\rho-1,7,10} + T^{2\rho-1,8,9})
\end{align*}

\item \cross{5}{I} gives
\begin{align*}
\boldsymbol{(\bar{3},2)(-1)(6)}: C^\nu_\rho = \epsilon_{\rho a b} \big[ &(T^{2a - 1,2b-1,2\nu+5} - T^{2a,2b,2\nu+5} - T^{2a - 1,2b,2\nu+6} - T^{2a,2b-1,2\nu+6}) + \\
&i(T^{2a - 1,2b-1,2\nu+6} - T^{2a,2b,2\nu+6} + T^{2a - 1,2b,2\nu+5} + T^{2a,2b-1,2\nu+5}) \big]
\end{align*}

\item \cross{5}{II} yields a singlet (the 8 vanishes)
\begin{align*}
\boldsymbol{(1,1)(-6)(6)}: C = \epsilon_{\beta a b} \big[ &(T^{2a - 1,2b-1,2\beta - 1} - T^{2a,2b,2\beta - 1} - 2T^{2a,2b-1,2\beta}) + \\
&i(T^{2a - 1,2b-1,2\beta} - T^{2a,2b,2\beta} + 2T^{2a,2b-1,2\beta-1}) \big]
\end{align*}

\item \cross{5}{III} gives
\begin{align*}
\boldsymbol{(\bar{3},2)(-7)(2)}: C^\nu_\rho = \epsilon^{\nu c} \epsilon_{\rho a b} \big[ &(T^{2a - 1,2b-1,2c+5} - T^{2a,2b,2c+5} + T^{2a - 1,2b,2c+6} + T^{2a,2b-1,2c+6}) + \\
&i(-T^{2a - 1,2b-1,2c+6} + T^{2a,2b,2c+6} + T^{2a - 1,2b,2c+5} + T^{2a,2b-1,2c+5}) \big]
\end{align*}

\item \cross{5}{IV} yields a $\bar{6}$, the 3 is not independent.
\begin{align*}
\boldsymbol{(\bar{6},1)(-2)(2)}: C_{\rho \sigma} = \epsilon_{\rho a b} \big[ &(T^{2a - 1,2b-1,2\sigma - 1} - T^{2a,2b,2\sigma - 1} + T^{2a - 1,2b,2\sigma} + T^{2a,2b-1,2\sigma}) + \\
&i(-T^{2a - 1,2b-1,2\sigma} + T^{2a,2b,2\sigma} + T^{2a - 1,2b,2\sigma - 1} + T^{2a,2b-1,2\sigma - 1}) \big]\\
+ &\{ \rho \leftrightarrow \sigma \}
\end{align*}

\item \cross{6}{I} gives
\begin{align*}
\boldsymbol{(3,2)(7)(-2)}: C^{\rho \nu} = \epsilon^{\rho a b} \big[ &(T^{2a - 1,2b-1,2\nu+5} - T^{2a,2b,2\nu+5} + T^{2a - 1,2b,2\nu+6} + T^{2a,2b-1,2\nu+6}) - \\
&i(-T^{2a - 1,2b-1,2\nu+6} + T^{2a,2b,2\nu+6} + T^{2a - 1,2b,2\nu+5} + T^{2a,2b-1,2\nu+5}) \big]
\end{align*}

\item \cross{6}{II} yields a $6$, the $\bar{3}$ is not independent.
\begin{align*}
\boldsymbol{(6,1)(2)(-2)}: C^{\rho \sigma} = \epsilon^{\rho a b} \big[ &(T^{2a - 1,2b-1,2\sigma - 1} - T^{2a,2b,2\sigma - 1} + T^{2a - 1,2b,2\sigma} + T^{2a,2b-1,2\sigma}) - \\
&i(-T^{2a - 1,2b-1,2\sigma} + T^{2a,2b,2\sigma} + T^{2a - 1,2b,2\sigma - 1} + T^{2a,2b-1,2\sigma - 1}) \big]\\
+ &\{ \rho \leftrightarrow \sigma \}
\end{align*}

\item \cross{6}{III} gives
\begin{align*}
\boldsymbol{(3,2)(1)(-6)}: C^{\rho \nu} = \epsilon^{\nu c} \epsilon^{\rho a b} \big[ &(T^{2a - 1,2b-1,2c+5} - T^{2a,2b,2c+5} - T^{2a - 1,2b,2c+6} - T^{2a,2b-1,2c+6}) - \\
&i(T^{2a - 1,2b-1,2c+6} - T^{2a,2b,2c+6} + T^{2a - 1,2b,2c+5} + T^{2a,2b-1,2c+5}) \big]
\end{align*}

\item \cross{6}{IV} yields a singlet (the 8 vanishes)
\begin{align*}
\boldsymbol{(1,1)(6)(-6)}: C = \epsilon^{\beta a b} \big[ &(T^{2a - 1,2b-1,2\beta - 1} - T^{2a,2b,2\beta - 1} - 2T^{2a,2b-1,2\beta}) - \\
&i(T^{2a - 1,2b-1,2\beta} - T^{2a,2b,2\beta} + 2T^{2a,2b-1,2\beta-1}) \big]
\end{align*}

\item \cross{9}{I} yields a (3,3) (the (3,1) is not independent)
\begin{align*}
\boldsymbol{(3,3)(-2)(2)}: C^{\rho \mu}_\nu = &(T^{2\rho-1,2\nu+5,2\mu+5} + T^{2\rho,2\nu+6,2\mu+5} - T^{2\rho,2\nu+5,2\mu+6} + T^{2\rho-1,2\nu+6,2\mu+6})+\\
&i(T^{2\rho-1,2\nu+5,2\mu+6} + T^{2\rho,2\nu+6,2\mu+6} + T^{2\rho,2\nu+5,2\mu+5} - T^{2\rho-1,2\nu+6,2\mu+5}) - \\
&\delta^\mu_\nu (T^{2\rho, 2\alpha + 6, 2\alpha + 5} - iT^{2\rho -1, 2\alpha + 6, 2\alpha + 5})
\end{align*}

\item \cross{10}{III} yields a $(\bar{3},3)$ (the $(\bar{3},1)$ is not independent)
\begin{align*}
\boldsymbol{(\bar{3},3)(2)(-2)}: C_{\rho \nu}^\mu = &(T^{2\rho-1,2\nu+5,2\mu+5} + T^{2\rho,2\nu+6,2\mu+5} - T^{2\rho,2\nu+5,2\mu+6} + T^{2\rho-1,2\nu+6,2\mu+6})-\\
&i(T^{2\rho-1,2\nu+5,2\mu+6} + T^{2\rho,2\nu+6,2\mu+6} + T^{2\rho,2\nu+5,2\mu+5} - T^{2\rho-1,2\nu+6,2\mu+5}) - \\
&\delta^\mu_\nu (T^{2\rho, 2\alpha + 6, 2\alpha + 5} + iT^{2\rho -1, 2\alpha + 6, 2\alpha + 5})
\end{align*}

\item \cross{10}{II} yields an (8,2) (the (1,2) is not independent)
\begin{align*}
\boldsymbol{(8,2)(3)(2)}: C^{\sigma \nu}_\rho = &(T^{2\rho-1,2\nu+5,2\sigma-1} + T^{2\rho,2\nu+6,2\sigma-1} + T^{2\rho,2\nu+5,2\sigma} - T^{2\rho-1,2\nu+6,2\sigma})+\\
&i(-T^{2\rho,2\nu+5,2\sigma-1} + T^{2\rho-1,2\nu+6,2\sigma-1} + T^{2\rho-1,2\nu+5,2\sigma} + T^{2\rho,2\nu+6,2\sigma}) - \\
&\sfrac{2}{3} \delta^\sigma_\rho (T^{2\beta, 2\nu + 6, 2\beta - 1} - iT^{2\beta, 2\nu + 5, 2\alpha -1})
\end{align*}

\item \cross{9}{IV} yields an (8,2) (the (1,2) is not independent)
\begin{align*}
\boldsymbol{(8,2)(-3)(-2)}: C^{\sigma \nu}_\rho = \epsilon^{\nu a} \big[&(T^{2\rho-1,2a+5,2\sigma-1} + T^{2\rho,2a+6,2\sigma-1} + T^{2\rho,2a+5,2\sigma} - T^{2\rho-1,2a+6,2\sigma})-\\
&i(-T^{2\rho,2a+5,2\sigma-1} + T^{2\rho-1,2a+6,2\sigma-1} + T^{2\rho-1,2a+5,2\sigma} + T^{2\rho,2a+6,2\sigma}) - \\
&\sfrac{2}{3} \delta^\sigma_\rho (T^{2\beta, 2a + 6, 2\beta - 1} + iT^{2\beta, 2a + 5, 2\alpha -1}) \big]
\end{align*}
\end{itemize}
Any other possible cross terms on the LHS are either zero or not independent of the ones listed above (due to antisymmetry of $T^{ijk}$). We now have the complete 120 breakdown:
\begin{equation} \label{eq:120breakdownappendix} 
\begin{array}{l}
120 = (45 \otimes 10)_A \longrightarrow \\ \\
\begin{array}{rccccccccc}
\Big[ & (1,1)(6)(-6) & \oplus & (1,1)(-6)(6) & \oplus & (1,2)(3)(2) & \oplus & (1,2)(-3)(-2) & \oplus\\
 & \scross{6}{IV} & & \scross{5}{II} & & \scross{1}{I} & & \scross{1}{III}\\ \\
 & (1,2)(3)(2) & \oplus & (1,2)(-3)(-2) & \oplus & (3,1)(-2)(2) & \oplus & (\bar{3},1)(2)(-2) & \oplus\\
 & \scross{2}{I} & & \scross{2}{III} & & \scross{1}{II} & & \scross{1}{IV}\\ \\
 & (3,1)(4)(6) & \oplus & (\bar{3},1)(-4)(-6) & \oplus & (3,1)(-2)(2) & \oplus & (\bar{3},1)(2)(-2) & \oplus\\
 & \scross{3}{II} & & \scross{4}{IV} & & \scross{2}{II} & & \scross{2}{IV}\\ \\
 & (3,1)(-8)(-2) & \oplus & (\bar{3},1)(8)(2) & \oplus & (3,2)(1)(-6) & \oplus & (\bar{3},2)(-1)(6) & \oplus\\
 & \scross{4}{II} & & \scross{3}{IV} & & \scross{6}{III} & & \scross{5}{I}\\ \\ 
 & (3,2)(7)(-2) & \oplus & (\bar{3},2)(-7)(2) & \oplus & (3,3)(-2)(2) & \oplus & (\bar{3},3)(2)(-2) & \oplus\\
 & \scross{6}{I} & & \scross{5}{III} & & \scross{9}{I} & & \scross{10}{III}\\ \\ 
 & (6,1)(2)(-2) & \oplus & (\bar{6},1)(-2)(2) & \oplus & (8,2)(3)(2) & \oplus & (8,2)(-3)(-2) & &\Big] \\
 & \scross{6}{II} & & \scross{5}{IV} & & \scross{10}{II} & & \scross{9}{IV} 
\end{array}
\end{array}
\end{equation}
(Instead of writing down explicitly how the representation is embedded under each term we give the reference to the appropriate entry in the list following Equation \ref{eq:120product}.)

\section{Bits and Pieces for the 16 calculation}
\label{ap:16}
The $M_S$ matrix is given by
\begin{equation} \label{eq:Sbasischange}
M_S = 
\left( \begin{array}{cccccccccccccccc}
	&&&&&&&&&&&&&&&-1 \\
	&&&&&&&&&&&&&&1& \\
	&&&&&&&&&&&&&-1&& \\
	&&&&&&&&&&&&1&&& \\
	&&&&&&&&&&&1&&&& \\
	&&&&&&&&&&-1&&&&& \\
	&&&&&&&&&1&&&&&& \\
	&&&&&&&&-1&&&&&&& \\
	&&&&&&&-1&&&&&&&& \\
	&&&&&&1&&&&&&&&& \\
	&&&&&-1&&&&&&&&&& \\
	&&&&1&&&&&&&&&&& \\
	&&&1&&&&&&&&&&&& \\
	&&-1&&&&&&&&&&&&& \\
	&1&&&&&&&&&&&&&& \\
	-1&&&&&&&&&&&&&&&
\end{array} \right)
\end{equation}

The couplings for eigenstates of the general Dirac Equation (\ref{eq:general16dirac}) are (up to a common factor of $\sfrac{1}{2}$)
\begin{align} \label{eq:funkygeneralfermioncouplings}
(1,1)(6)(-1): \qquad  &  4a_{min}(g_A - g_B) + m_A + m_B + 6 a_{min} (g_A - g_B) \tanh(z \mu) \pm \nonumber \\
& (4m_{AB}^2 + (4a_{min}(g_A + g_B) + m_A - m_B + 6 a_{min} (g_A + g_B)\tanh(z \mu))^2)^{\frac{1}{2}} \nonumber \\
& \nonumber \\
(1,1)(0)(-5): \qquad   &  4a_{min}(-g_A + g_B) + m_A + m_B + 6 a_{min} (g_A - g_B) \tanh(z \mu) \pm \nonumber \\
& (4m_{AB}^2 + (4a_{min}(g_A + g_B) - m_A + m_B - 6 a_{min} (g_A + g_B)\tanh(z \mu))^2)^{\frac{1}{2}} \nonumber \\
& \nonumber \\
(1,2)(-3)(3): \qquad   &m_A + m_B + 6 a_{min} (-g_A + g_B) \tanh(z \mu) \pm \nonumber \\
& (4m_{AB}^2 + (-m_A + m_B + 6 a_{min} (g_A + g_B)\tanh(z \mu))^2)^{\frac{1}{2}}   \nonumber \\
& \nonumber \\
(\bar{3},1)(-4)(-1):  \qquad  & 4a_{min}(-g_A + g_B) + m_A + m_B + 2 a_{min} (-g_A + g_B) \tanh(z \mu) \pm \nonumber \\
& (4m_{AB}^2 + (4a_{min}(g_A + g_B) - m_A + m_B + 2 a_{min} (g_A + g_B)\tanh(z \mu))^2)^{\frac{1}{2}} \nonumber \\
& \nonumber \\
(\bar{3},1)(2)(3):   \qquad &  4a_{min}(g_A - g_B) + m_A + m_B + 2 a_{min} (-g_A + g_B) \tanh(z \mu) \pm \nonumber \\
& (4m_{AB}^2 + (4a_{min}(g_A + g_B) + m_A - m_B -  2 a_{min} (g_A + g_B)\tanh(z \mu))^2)^{\frac{1}{2}} \nonumber \\ 
& \nonumber \\
(3,2)(1)(-1):  \qquad  &  m_A + m_B + 2 a_{min} (g_A - g_B) \tanh(z \mu) \nonumber \\
& \pm (4m_{AB}^2 + (m_A - m_B + 2 a_{min} (g_A + g_B)\tanh(z \mu))^2)^{\frac{1}{2}}
\end{align}

\bibliography{thesis}

\end{document}